\documentclass[%
 reprint,
superscriptaddress,
 amsmath,amssymb,
 aps,
twocolumn]{revtex4-2}

\usepackage{graphicx}% Include figure files
\usepackage{dcolumn}% Align table columns on decimal point

\usepackage[dvipsnames]{xcolor}
\usepackage{stackrel}
\usepackage{siunitx}
\usepackage{comment}

\usepackage[english]{babel}
\makeatletter
\@namedef{l@en}{\l@english}
\@namedef{captionsen}{\captionsenglish}
\@namedef{daten}{\dateenglish}
\makeatother

\usepackage{amsfonts}

\usepackage{bbold}

\usepackage{hyperref}
\usepackage[shortlabels]{enumitem}

\usepackage{multirow}

\usepackage{lineno}
\newcommand{\brak}[1]{\langle #1 \rangle}
\newcommand{\Nmean}{\brak{N}}
\newcommand{\dd}{\mathrm{d}}
\newcommand{\kk}{\boldsymbol{k}}
\newcommand{\K}{\mathop{\mathrm{K}}\displaylimits}
\newcommand{\bmr}{\boldsymbol{r}}
\newcommand{\vaoup}{\boldsymbol{v}_{\rm OU}}
\newcommand{\dotvaoup}{\dot{\boldsymbol{v}}_{\rm OU}}

\AtBeginDocument{%
  \setlength{\linenumbersep}{\dimexpr\oddsidemargin+0.48in-\linenumberwidth\relax}%
}

\begin{document}
%\linenumbers

\preprint{APS/123-QED}

\title{The Countoscope for self-propelled particles}

\author{Tristan Cerdin}%
\affiliation{Université Paris-Saclay, CNRS, FAST, 91405, Orsay, France}
\affiliation{%
CNRS, Sorbonne Université, Physicochimie des Electrolytes et Nanosystèmes Interfaciaux, F-75005 Paris, France
}%

\author{Talia Calazans}%
\affiliation{Department of Physics and astronomy, University of Pennsylvania, Philadelphia, USA
}%

\author{C. Douarche}
\affiliation{Université Paris-Saclay, CNRS, FAST, 91405, Orsay, France}

\author{Sophie Marbach}%
\affiliation{%
CNRS, Sorbonne Université, Physicochimie des Electrolytes et Nanosystèmes Interfaciaux, F-75005 Paris, France
}%
\email{sophie.marbach@cnrs.fr}

\begin{abstract}
Particle number fluctuations $N(t)$, measured in virtual observation boxes of an image or a simulation, offer a way to quantify particle dynamics when particle tracking is impractical, such as in high-density systems. While traditionally limited to equilibrium diffusive systems, we extend this approach -- named ``Countoscope'' -- to out-of-equilibrium self-propelled particles: Active Brownian (ABPs), Run and Tumble (RTPs), and Active Ornstein-Uhlenbeck Particles (AOUPs). For AOUPs, we leverage their Gaussian statistics to derive a general formula applicable to any Gaussian system. For ABPs and RTPs, we derive the intermediate scattering function (ISF)—and thus the correlations of $N(t)$—using an exact perturbative expansion over the probability density fields, revealing key physical features of the ISF and of the number correlations. Our theoretical predictions for the mean-squared number difference $\langle \Delta N^2(t) \rangle = \langle (N(t) - N(0))^2 \rangle$ match stochastic simulations and exhibit three time-dependent scaling regimes: diffusive, advective, and long-time enhanced diffusive, reflecting the regimes of the mean squared particle displacement. We further uncover limiting laws in each of these regimes that are useful to quantify self-propulsion properties.
\end{abstract}

\maketitle

More than a century ago, Smoluchowski introduced the concept of measuring particle dynamics by probing the number \( N(t) \) of particles within a microscope's field of view~\cite{smoluchowski1914studien,smoluchowski1915uber,smoluchowski1916studien,chandrasekhar1943stochastic}. As particles undergo Brownian motion, they diffuse in and out of the observation region, inducing fluctuations in \( N(t) \). Smoluchowski developed a theory -- later verified experimentally~\cite{westgren1916veranderungsgeschwindigkeit,westgren1918veranderungsgeschwindigkeit} -- to relate diffusion coefficients to the correlations of \( N(t) \) at equilibrium. While a limited number of studies expanded on this principle through the 1980s, most remained theoretical, with even fewer addressing out-of-equilibrium systems~\cite{lindley1954estimation,rothschild1953new,ruben1964generalised,katz1975methods,culling1985estimation}. A notable 1953 study by Rothschild examined the dynamics of sperm cells~\cite{rothschild1953new}, a canonical example of self-propelling particles that consume energy to perform directed motion~\cite{vrugt2025exactly}. Assuming sperm cells swim straight without reorientation, Rothschild derived a theory to quantify their swim speed from fluctuating counts. He also introduced the idea of counting particles in many virtual observation boxes of an image to improve statistical accuracy. However, due to the practical challenges of manual counting and correlation analysis at the time, the potential of number fluctuations was progressively forgotten~\cite{baker1957spermatozoan}.

Smoluchowski's modern legacy lies instead in techniques such as Dynamic Light Scattering (DLS) and Fluorescence Correlation Spectroscopy (FCS), which rely on correlations of the \textit{intensity} scattered by particles within an illuminated region~\cite{berne2000dynamic,elson1974fluorescence}. These methods are typically used to measure diffusion coefficients in equilibrium suspensions and are less commonly applied to out-of-equilibrium systems. Since self-propelled particles exhibit both directed motion and reorientation over time, their long-time behavior is akin to enhanced diffusion. While DLS and FCS can quantify this enhanced diffusion~\cite{gunther2018diffusion,santiago2018nanoscale}, and ongoing research continues to refine their ability to distinguish self-propulsion speed and reorientation time from the signals~\cite{mcglasson2021investigating}.

Advances in modern microscopy and automated analysis have removed the obstacles that once made Smoluchowski's method impractical. Recent work by some of us has revived his approach to investigate the diffusive properties of colloidal suspensions at equilibrium~\cite{mackay2024countoscope}. Importantly, we introduced the ``Countoscope'' concept, which involves probing \( N(t) \) in observation boxes of varying sizes to resolve dynamics across different length scales. This approach has enabled us to identify a peculiar hydrodynamic enhancement of \textit{collective} diffusion in dense colloidal suspensions~\cite{carter2025measuring}. However, the potential of this method for analyzing out-of-equilibrium systems -- particularly self-propelled particles -- remains unexplored.

The primary challenge in applying Smoluchowski's formalism to self-propelled particles is the absence of a theoretical framework linking self-propelled dynamical parameters to fluctuating counts. Since \( N(t) \) can be expressed as the integral of particle density within a box, a model for density correlations is sufficient to infer the correlation functions of \( N(t) \). The intermediate scattering function (ISF), defined as the spatial Fourier transform of the density correlation, is a standard tool for investigating such correlations. Recent efforts have focused on solving the exact ISF for non-interacting self-propelled systems~\cite{martens_probability_2012,kurzthaler_intermediate_2016,kurzthaler2024characterization,zhao_quantitative_2024}, achieving remarkable agreement with experimental data from Janus colloids and bacteria~\cite{kurzthaler_intermediate_2016,kurzthaler2024characterization}. While some of these approaches are model-specific~\cite{martens_probability_2012} or rely on sophisticated theoretical frameworks requiring numerical solutions to obtain the ISF~\cite{kurzthaler_intermediate_2016,zhao_quantitative_2024}, they highlight the rich potential of ISFs to investigate self-propelled particles.

% However, these approaches are either model-specific~\cite{martens_probability_2012} or rely on complex theoretical frameworks requiring numerical solutions to obtain the ISF~\cite{kurzthaler_intermediate_2016,zhao_quantitative_2024}. This limits the practical utility of existing exact ISF approaches for analyzing number fluctuations. New formulations of the ISF for self-propelled particles are therefore needed.

To investigate number correlations with ISFs, it is useful to manipulate simple expressions of the ISFs, to increase computational efficiency as number correlations require additional integration steps. 
Recently, several theories have been developed to solve the probability distribution functions of self-propelled particles through the Fokker-Planck equation (FPE) of the system~\cite{martin_statistical_2021,dinelli_fluctuating_2024,gautry_closures_2025,solon2015active}. 
These approaches, while perturbative, offer simplified solutions and, therefore, hold significant potential for deriving analytical laws for number fluctuations. Such a perturbative strategy on the FPE has recently been successfully applied to obtain a simplified ISF of anisotropic Active Brownian Particles~\cite{gautry_closures_2025}. Another advantage is that these theories allow different models of self-propelled particles to be treated within the same framework~\cite{dinelli_fluctuating_2024,solon2015active}, making them highly versatile.

%ypically  only the long time diffusion coefficient. 
%\cite{santiago2018nanoscale} typically more for nanoscale, and also just long time enhanced collective diffusion. and this one for DLS goes a bit further but typically shows you can not uncouple the different dynamical parameters. 

% a century ago, a method in real space has recently been developed~\cite{mackay2024countoscope} that relies on the fluctuation of the number of particles in virtual observation boxes in time to quantify particle dynamics. Of the spectral methods only DDM has been widely applied to the study of self-propelling systems through the achievement of complex theoretical work.\tristan{Is this true ? I may be lying.} No real space equivalent exists yet in this field, and this paper aims to bridge this gap by establishing the theoretical framework necessary for the use of the Countoscope method to quantify the motion of active particle systems.

% None of these existing approach to obtain ISFs have been used to investigate number theories that can complement them by offering a possibility for analysis of dynamics in real space from simple microscopy data.   

Here, we derive theoretical expressions for the number fluctuations of three canonical models of self-propelled non-interacting particles: Active Brownian Particles (ABP), Run-and-Tumble Particles (RTP), and Active Ornstein-Uhlenbeck Particles (AOUP). We use a perturbative expansion of their Fokker-Planck equations, conducted over increasing modes of orientational order. This expansion can also be interpreted as a perturbation in the P\'eclet number governing orientational relaxation. Our approach to obtain the Fourier-time ISF goes beyond prior works since we obtain all terms of the expansion, producing exact results in the form of an infinite series of fractions -- a so called \textit{continued} fraction. The real-time ISF is obtained via inverse Fourier transform, with explicit expressions available at select perturbation levels. Number correlations are then derived as spatial integrals of the ISF.
Comparison with numerical simulations confirms that our expressions are exact, with accuracy maintained for sufficiently small P\'eclet numbers when truncated at finite perturbation orders. 

Our theoretical expansion also clarifies key features of the ISF and number correlations: oscillations in both correlation functions arise from reorientation dynamics. Importantly, number correlations are very sensitive to differences in reorientational behavior between the different models. We identify three dynamical regimes in the number correlations: translational diffusion at short times, self-propulsion at intermediate times, and enhanced diffusion at long times. These limiting regimes reflect behavior of the mean squared displacement of particles, and are each more pronounced in small or large observation boxes depending on the regime. Finally, we obtain simple theoretical laws for number correlations in each of these regimes and discuss them in the practical context of extracting dynamical parameters from experimental data.

\section{Setup for Counting Self-Propelled Particles}

\subsection{Counting Method}
\label{sec:counting intro}
\begin{figure}[h!]
    \centering
    \includegraphics[width=0.99\linewidth]{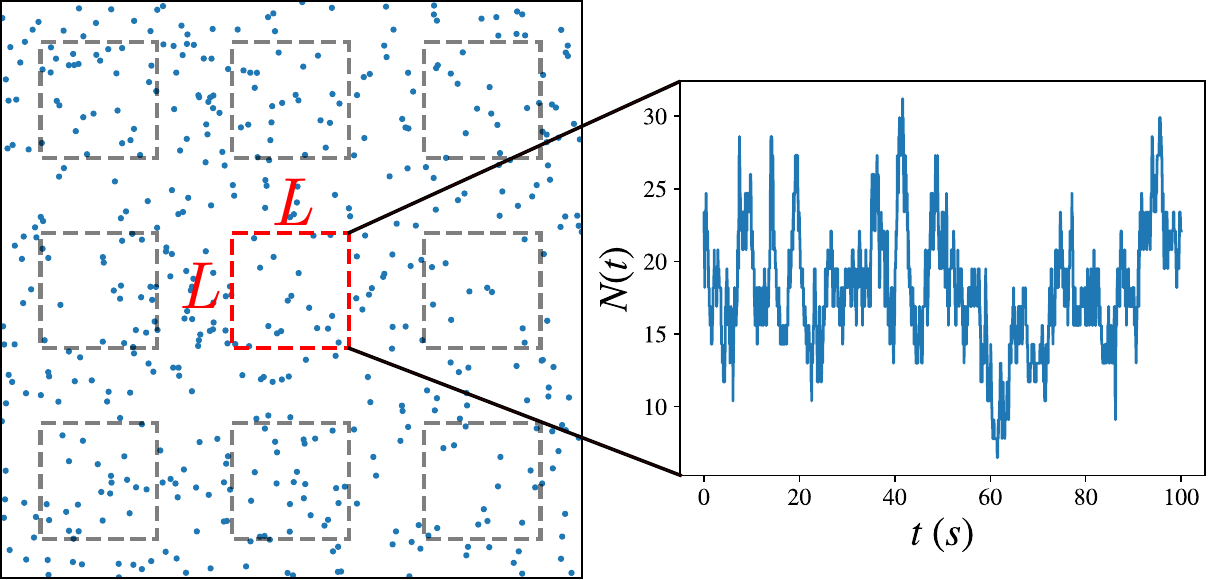}
    \caption{\textbf{Probing number fluctuations} ``Countoscope'' approach where we probe the number of particles in virtual observation boxes of an image/simulation. }
    \label{fig:fig1}
\end{figure}

We study the number $N(t)$ of self-propelled particles in square boxes of size $L\times L$. Note, that although we provide a reasoning here for two dimensions, the tools we use may readily be adapted to three dimensions~\cite{gautry_closures_2025}. The number $N(t)$ fluctuates as particles randomly enter and exit the box, undergoing various dynamics, see an example in Fig.~\ref{fig:fig1}. To read out the particles' dynamic properties from the fluctuating counts, we will here mostly explore 2 correlation functions: (i) the time correlation function $C_N(t) = \brak{N(t) N(0)} - \Nmean^2$ for the particle number and (ii) the mean square change in particle number
\begin{equation}
    \begin{split}
\label{eq:counting}
    \langle \Delta N^2(t)\rangle  &= \brak{\left(N(t) - N(0)\right)^2} \\
    &= 2 (\brak{N^2} - \Nmean^2) - 2 (\brak{N(t) N(0)} - \Nmean^2) \\
   &=  2 (\brak{N^2} - \Nmean^2) - 2 C_N(t).
\end{split}
\end{equation}
In the above, $\langle \cdot \rangle$ corresponds to an average over realizations of the noise, initial conditions and boxes of same size in an image. 
While we do not explicitly highlight this in our notations signatures, we note that $N(t)$ crucially depends on the size of the virtual box, and probes different motion occurring at different length scales $L$. 

$\langle \Delta N^2(t)\rangle $ can be put in parallel, for particle numbers, to the Mean Squared Displacement (MSD) $\langle \Delta \boldsymbol{r}^2(t) \rangle = \langle |\boldsymbol{r}(t) - \boldsymbol{r}(0)|^2\rangle$, where $\boldsymbol{r}(t)$ refers to a particle's position at time $t$. We will refer throughout this article to $\langle \Delta N^2(t)\rangle$ as the Number Mean Squared Difference (NMSD).
One main difference is that whereas for the MSD the averaging $\langle \cdot\rangle$ is done over individual particle trajectories, for the NMSD the averaging is instead taken over all boxes $L$ ``drawn'' over the observation window. Note, that while they don't in Fig.~\ref{fig:fig1}, typically observation boxes overlap to allow for increased statistics, as demonstrated in a previous work~\cite{carter2025measuring}. Typically, here the number of boxes ranges between $10^1-10^3$ for each box size, with fewer boxes for bigger box sizes.

In our investigation, we will consider for simplicity that particles do not interact. The presence or not of a particle in a box at fixed time $t$ is therefore independent of that of others and, therefore, follows a Poisson process. Hence, we can write $\langle N^2 \rangle - \langle N \rangle^2 = \langle N \rangle$~\cite{nelson2008penguin}. We can thus simplify Eq.~\eqref{eq:counting} to $\langle \Delta N^2(t)\rangle  = 2\Nmean - 2C_N(t)$. It is thus sufficient to obtain a theoretical expression for either $C_N(t)$ or the NMSD to obtain predictions on both quantities.   
Here we will focus on $C_N(t) = \langle N(t)N(0) \rangle - \langle N\rangle^2 $. We can write $N(t)$ in terms of individual particle contributions $N(t) = \sum_i n_i(t)$, where $n_i = 1 $ when particle $i$ is inside the box and $0$ when outside~\cite{mackay2025collective}. 
Then for non-interacting particles
\begin{align*}
    \langle N(t) N(0) \rangle &= \sum_i \langle n_i(t) n_i(0) \rangle + \sum_{i, j\neq i} \langle n_j(t)n_i(0)\rangle \\
    &=\langle N \rangle P_{\mathrm{in}}(t) + \langle N \rangle^2
\end{align*}
where $P_{\mathrm{in}}$ is the probability that a particle present in the box at $t=0$ is still in it at time $t$ ($n_i(t) = n_i(0) = 1$).
$C_N(t)$ can thus be re-expressed as 
\begin{equation}
\label{eq:probability}
    C_N(t) = \langle N \rangle P_{\mathrm{in}}(t)
\end{equation}
 Note, that this relationship was established quite some time ago by first principles by Smoluchowski~\cite{smoluchowski1914studien,smoluchowski1916studien}. The factor $P_{\mathrm{in}}(t)$ was then referred to as ``probability after-effect''~\cite{chandrasekhar1949}. While its expression for strictly diffusive systems is known, it is much less intuitive for self-propelling particles. Overall, $\langle N(t) N(0) \rangle$ has an $\langle N \rangle^2$ contribution from different particles, and an excess contribution from how long each of the $\Nmean$ particles initially inside the box keeps contributing to correlations by staying in the box~\cite{mackay2025collective}. 

We can further express $P_{\mathrm{in}}(t)$ in terms of $a_0(\boldsymbol{r},\boldsymbol{r'},t)$, the probability of a particle to be at position $\boldsymbol{r}$ at time $t$ with initial position $\boldsymbol{r'}$. We do so by integrating on $\boldsymbol{r}$ over the size of the box and averaging over all possible starting positions $\boldsymbol{r'}$ within the box. %as 
% \begin{equation}
% \begin{split}
%     P_{\mathrm{in}}(t) &= \frac{1}{L^2}\iint_{[-\frac{L}{2},\frac{L}{2}]^2}d\boldsymbol{r}d\boldsymbol{r'} \, a_0(\boldsymbol{r},\boldsymbol{r'},t) \\
%     &= \frac{1}{L^2}\iint_{[-\frac{L}{2},\frac{L}{2}]^2}d\boldsymbol{r}d\boldsymbol{r'} \, a_0(\boldsymbol{r}-\boldsymbol{r'},t) 
% \end{split}
% \end{equation}We do it by integrating over the size of the box that the particle has to reach over $\boldsymbol{r}$, and taking an average all possible the initial conditions $\boldsymbol{r'}$ inside the box. 
% Given that $a_0$ depends only on the distance between $\mathrm{r}$ and $\mathrm{r'}$
Rewriting $a_0$ as $a_0(\boldsymbol{r}-\boldsymbol{r'},t)$, since $a_0$ should only depend on the distance between start and end point due to translational invariance, we obtain
\begin{equation}
    P_{in}(t) = \frac{1}{L^2}\iint_{[-\frac{L}{2},\frac{L}{2}]^2}d\boldsymbol{r}d\boldsymbol{r'} \, a_0(\boldsymbol{r}-\boldsymbol{r'},t).
\end{equation}

It will often be useful to operate in spatial Fourier space, and so we define  $F(\boldsymbol{k},t)$, the Fourier transform of $a_0$ as
\begin{equation}
\label{eq:ISF def}
     F(\boldsymbol{k},t) = \int d\boldsymbol{r} e^{-i\boldsymbol{k}\cdot \boldsymbol{r}}a_0(\boldsymbol{r},t)
\end{equation}
for a given wavenumber $\kk$. $F(\boldsymbol{k},t)$ is the self Intermediate Scattering Function (ISF) of the particles When particles are non-interacting, and writing $\boldsymbol{r'}$ their  initial position, $F(\boldsymbol{k},t)$ can be written~\cite{berne2000dynamic} as :
\begin{equation}
    \label{eq:ISF nonin}
    F(\boldsymbol{k},t) \equiv  \langle e^{-i\boldsymbol{k}\cdot(\boldsymbol{r} - \boldsymbol{r'})} \rangle = \int d(\boldsymbol{r}-\boldsymbol{r'})e^{-i\boldsymbol{k}\cdot(\boldsymbol{r}-\boldsymbol{r'})} a_0(\boldsymbol{r}-\boldsymbol{r'},t)
\end{equation}
where the average indicates an average over all particle trajectories. This last formulation shows that $F(k,t)$ is indeed the Fourier transform of $a_0$. We thus obtain 
% We can then easily calculate $F(\boldsymbol{k},t) $ from our expressions for $\widetilde{a}_0(\boldsymbol{k},\omega)$ by simply inverting the Fourier transform on time.
\begin{equation}
    \label{eq:Cnt}
    C_N(t)= \frac{\langle N \rangle}{L^2}\iint_{[-\frac{L}{2},\frac{L}{2}]^2} d\boldsymbol{r}d\boldsymbol{r'} \int \frac{d\boldsymbol{k}}{(2\pi)^2} e^{i\boldsymbol{k}\cdot(\boldsymbol{r}-\boldsymbol{r'})} F(\boldsymbol{k},t).
\end{equation}
Note that this expression can be simplified by carrying out the integrals on $\boldsymbol{k}$ -- see Eq.~\eqref{eq:c_n k int}. In this paper, we will establish expressions for either directly $P_{\rm in}(t)$ or $F(\boldsymbol{k},t)$ to obtain expressions for the correlation function $C_N(t)$, through Eq.~\eqref{eq:probability} or Eq.~\eqref{eq:Cnt}.

% It is formally defined as :
% \begin{equation}
% \label{eq:ISF def}
%     F(\boldsymbol{k},t) = \frac{1}{N}\langle \widetilde{\hat{\rho}}(\boldsymbol{k},t) \widetilde{\hat{\rho}}^*(\boldsymbol{k},0) \rangle
% \end{equation}
% With $\hat{\rho}(\boldsymbol{r},t) = \sum_i^N \delta(\boldsymbol{r} - \boldsymbol{r_i}) $ the local density of particles.

\subsection{Three models for self-propelled particles}
\label{sec:active models}

We consider 3 broadspread models of self-propelled particles to benchmark the ``Countoscope'' approach in this nonequilibrium context. First, we consider Run-and-Tumble Particles (RTP), which describe well a range of biological systems, including many bacteria and swimming microalgae~\cite{berg_e_2004, kurzthaler2024characterization, clement_bacterial_2016, polin2009chlamydomonas}. In its simplest form, in 2D, the position $\boldsymbol{r}(t)$ of a particle whose orientation is given by $\boldsymbol{u}(\theta) = (\cos{\theta}, \sin{\theta})$ evolves as
\begin{align}
    &\dot{\boldsymbol{r}}(t) = v\boldsymbol{u}(\theta_i) + \sqrt{2D_t} \boldsymbol{\eta}(t) \label{eq:roft}\\
    & \theta \underset{\alpha}{\rightarrow} \theta' \in [0,2\pi] \label{eq:theta}
\end{align}
where $v$ is the self-propulsion speed, $D_t$ a translational diffusion coefficient, and $\boldsymbol{\eta}(t)$ a Gaussian white noise satisfying $\langle \boldsymbol{\eta}(t)\rangle=0$ and $\langle \eta_i(t)\eta_j(t')\rangle= \delta_{ij}\delta(t-t')$ where $\langle \cdot \rangle$ represents an average over realizations of the noise. Eq.~\eqref{eq:theta} corresponds to exponentially distributed tumbling events at rate $\alpha$. Note, that some models include 2 rates to distinguish tumbling and running phases, but in the vast majority of cases, one can take the approximation of instantaneous tumbles~\cite{berg_e_2004}. Second, we consider the Active Brownian Particle (ABP) model, which describes better systems where reorientation is smooth and diffusive. In that case the evolution equation for $\boldsymbol{r}(t)$ is the same as for the RTP Eq.~\eqref{eq:roft}, but the orientation evolves as
\begin{equation}
    \dot{\theta}(t) = \sqrt{2D_r} \eta_r(t)
\end{equation}
where $\eta_r(t)$ is another Gaussian white noise and $D_r$ is the angular diffusion coefficient. ABP models capture well the motion of Janus particles~\cite{kurzthaler2018probing, poncet_pair_2021,sprenger_active_2020, fernandez-rodriguez_feedback-controlled_2020}. For both RTP and ABP, fluctuations come from passive diffusivity and reorientation.
It can however be useful to allow the self propulsion speed to fluctuate for certain applications, which is captured by the Active Ornstein Uhlenbeck Particles model (AOUP). In this case, %\tristan{rename $\boldsymbol{v}$ into $\vaoup$ (reverbarate in PRL as well)}
\begin{align}
    &\dot{\boldsymbol{r}}=\vaoup(t)+\sqrt{2D_t}\boldsymbol{\eta}(t) \\
    &\tau_r\dotvaoup= -\vaoup(t) + \sqrt{2D_v}\boldsymbol{\eta}_v(t)
\end{align}
where $\tau_r$ is a timescale associated with velocity relaxation, $D_v$ a velocity diffusion coefficient and $\boldsymbol{\eta}_v(t)$ another Gaussian random noise. In this model the mean particle velocity is $v \equiv \langle \sqrt{|\vaoup^2(t)|} \rangle = \sqrt{2D_v/\tau_r}$, allowing us to build correspondence between the different models.

For all three models, the MSD can be captured by a single equation
\begin{equation}
    \label{eq:MSD}
    \langle \Delta \boldsymbol{r}^2(t) \rangle  = 4 \left(D_t +\frac{v^2}{2D_r}\right)t + 2\frac{v^2}{D_r^2}(e^{-D_rt} -1 )
\end{equation}
where $D_r = \alpha $ for RTPs, and $D_r = 1/\tau_r$ and $v = \sqrt{2D_v/\tau_r}$ for AOUPs. 
This equation captures the 3 regimes of motion displayed by the particles: at very short times, motion is diffusive $\langle \Delta \bm{r}^2(t) \rangle = 4D_t t$ until a crossover time $t_{adv} = 4D_t/v^2$, after which motion is ballistic $\langle \Delta \bm{r}^2(t) \rangle  = v^2 t^2$. Reorientation dynamics make motion diffusive again after another crossover time $t_{diff} = 4 D_{\rm eff}/v^2$ and $\langle \Delta \bm{r}^2(t) \rangle = 4D_{\rm eff}t$  with an effective diffusion coefficient  $D_{\rm eff} = D_t +\dfrac{v^2}{2D_r}$ that is enhanced by activity.  We expect these three regimes to also appear in the NMSD due to the similarities in how they probe particle dynamics in time.

We quantify the impact of activity on the motion of active particles with a dimensionless P\'eclet number $ \text{Pe}$,
\begin{equation}
    \label{eq:Péclet}
    \text{Pe} = \frac{v\ell}{D_t} = \frac{v}{2\sqrt{D_r^\alpha D_t}}.
\end{equation}
 where $D_r^\alpha = D_r$ for ABP and $\alpha$ for RTP. For convenience, we use the equivalent notation $D_r^\alpha = D_r +\alpha$. The $\text{Pe}$ number compares the timescale of active advection to that of diffusion, over a length scale $\ell$. We choose $\ell$ to characterize the distance that a particle diffuses before significant reorientation happens, so such that $4\ell^2/D_t = 1/D^\alpha_r$, giving $\ell = \sqrt{D_t/D_r^\alpha}/2$.
The synthetic Janus particles typically described by the ABP model tend to have a low $\text{Pe} \simeq 3-6$~\cite{sprenger_active_2020, kurzthaler2018probing,howse_self-motile_2007}. Bacteria such as \textit{E. coli} usually have higher self propulsion speed with $\text{Pe} \simeq 15-20$~\cite{zhao_quantitative_2024, berg_e_2004}.  

All theoretical calculations are checked against numerical simulations of the three particle models. For more details on numerical implementation, see Appendix~\ref{appendix:computational methods}.

\subsection{Calculating ISFs for self-propelled particles}

\paragraph{Non-gaussian models.}
For both the non interacting ABP and RTP models -- which are non-Gaussian models -- exact results of ISFs have been derived in recent years \cite{martens_probability_2012,kurzthaler_intermediate_2016,kurzthaler2024characterization}. However, these solutions are either model specific or computationally expensive. Given that investigating number fluctuations requires to perform an integral on $\boldsymbol{k}$ on top of the ISF, as given in Eq.~\eqref{eq:Cnt}, we attempt to find a numerically efficient expression of the ISF. In addition, given the complexity of the previously obtained expressions, it is hard to infer behavior in limiting regimes of parameters. 

We note that one common alternative approach to obtain ISFs involves stochastic density field theories  (sDFT), or Dean-Kawasaki equations~\cite{mackay2024countoscope,minh2023ionic,demery2016conductivity,jardat2022diffusion,mahdisoltani2021transient,minh2023ionic,bernard2023analytical,illien2025dean}. For this application, stochastic density field theory is impractical as there is no consensus yet on a stochastic density field the  ory for self-propelled particles~\cite{kuroda_microscopic_2023,das_coarse-grained_2013,duran-olivencia_general_2017,yadav_hydrodynamic_2022}. Rather, we will use approaches based on the FPE -- for which a consensus exists on the expression of the FPE for non interacting suspensions of active particles. Thus we obtain results directly on the probability distribution $a_0(\boldsymbol{r},t)$ rather than on the microscopic density $\sum_i \delta(\boldsymbol{r} - \boldsymbol{r}_i(t))$ studied by sDFT. %\tristan{I tried to keep this short Carine tell me if it's clearer or if it makes you more confused, this is linked to another comment we make later, idk if this should move}
%We will thus seek here computationally more efficient solutions for the ISFs.  
%To solve for ABPs one needs to numerically solve an eigenvalue problem, while the RTP solutions requires either a numeric inverse Laplace Transform on time or solving a complex system of equations to obtain results for $F(\boldsymbol{k},t)$.

\paragraph{Gaussian models.}
The case of AOUPs is slightly different as it is a Gaussian model. Indeed, the self propulsion speed $\vaoup(t)$ in AOUPs evolves smoothly as a Gaussian colored noise. Thus, the probability distribution of the position of non interactive AOUPs is Gaussian. It therefore satisfies the common property that all of the cumulants of its position variables above 2 are equal to 0. %All moments above the MSD are thus vanishing and we can easily compute the ISF. 
Looking back on Eq.~\ref{eq:ISF nonin} with initial condition $\boldsymbol{r}(0)$ we can show through a cumulant expansion~\cite{kubo_generalized_1962} that 
\begin{align*}
    F(\boldsymbol{k},t) = \langle e^{-i\boldsymbol{k} \cdot(\boldsymbol{r}(t)-\boldsymbol{r}(0))} \rangle &= e^{-i\boldsymbol{k} \cdot \langle \boldsymbol{r} - \boldsymbol{r}(0)) \rangle - \frac{k^2}{4} \langle |\boldsymbol{r}(t) - \boldsymbol{r}(0) |^2 \rangle +...} \\ &= e^{-\frac{k^2}{4}\langle \Delta r^2(t) \rangle},
\end{align*}
obtaining a well-known result for AOUPs using Eq.~\eqref{eq:MSD}.

This result can further be plugged in Eq.~\eqref{eq:Cnt}. Then, one can split up integrals on $x$ and $y$, and realizing they are the same, obtain 
%C_N(t) &= \langle N \rangle \left[2 \int \frac{du}{2\pi} \left( \frac{\sin(u)}{u} \right)^2 e^{-\frac{u^2}{L^2} \langle \Delta r^2(t) \rangle} \right ]^2 \\
\begin{align}
    C_N(t) &= \langle N \rangle \left[ f\left( \frac{\langle \Delta r^2(t) \rangle}{L^2} \right) \right]^2 \label{eq:Cnt Gaussian} \\
     & \text{with} \,\, f(\tau) = \left[ \sqrt{\frac{\tau}{\pi}} (e^{-1/\tau} -1) + \text{erf}\left(\sqrt{\frac{1}{\tau}}\right)\right] \notag
\end{align}
where $\text{erf}$ is the error function.  For more details on this calculation we refer the reader to Ref.~\cite{mackay2024countoscope}. 

It then follows that Countoscope formulas for \textit{any} Gaussian model can easily be obtained so long as one knows the MSD of the model, and simply by plugging the MSD into the function $f$. Remarkably the intuition for this formula was brought forwards recently in Ref.~\cite{metzler2025discriminating}, but not derived. Note that in Ref.~\cite{metzler2025discriminating} it was applied to both gaussian (Fractional Brownian motion) and non gaussian random walks (obstructed diffusion) with only a mild deviation observed for non gaussian cases. Eq.~\eqref{eq:Cnt Gaussian} will be useful as a reference point for the study of ABPs and RTPs, to see which role non gaussianity plays in number fluctuations.

\section{Hydrodynamic equations of motion for non gaussian systems}

\subsection{Derivation of the probability densities}
\label{sec:hydro demonstration}

%In order to obtain the NMSD we're interested in, we need an expression for the density of our active particles. To obtain it, we now turn to the use of a coarse graining framework to obtain hydrodynamics equations for the relevant macroscopic fields. We start from the stochastic equations of motions of our particles and use the standard procedure to obtain a Fokker-Planck equation for the probability density of a particle to be at position $\boldsymbol{r}$ with orientation $\theta$ at time $t$. This is to be preferred to the alternative coarse-graining method that is used in the literature which employs Stochastic Density Field Theory to directly derive evolution equations for the density $\hat{\rho}(\boldsymbol{r},t) = \sum_{i=1}^N \delta(\boldsymbol{r}-\boldsymbol{r}_i) $ [CITE ARTICLES]. This density $\hat{\rho}$ is still a microscopic object defined as a sum of delta functions and isn't smooth at large scales, and therefore it cannot be considered fully coarse grained yet. For active matter systems this can lead one to ignore relevant terms in the expansion when one treats it as a macroscopic variable, especially since an universal step in calculations using this method is to linearize this density around an average one $\rho_0$ which is ill defined considering the microscopic nature of this object. [CITE]
\paragraph{Formulation in terms of an infinite hierarchy.} To obtain expressions for the ISFs of our models we will rely on a coarse graining procedure going from the stochastic equations of motions to probability space. Using a Fokker-Planck equation (FPE) formalism we study the time evolution of $P(\boldsymbol{r},\theta,t)$, the probability density that a particle is found at position $\boldsymbol{r}$ with orientation $\theta$ at time $t$. 
For either an ABP or an RTP,
\begin{equation}
    \begin{split}
        \partial_tP(\boldsymbol{r},\theta,t)=-\nabla_r \cdot \left(v\boldsymbol{u}(\theta)P - D_t\nabla_rP\right) + D_r\partial^2_\theta P \\- \alpha P + \frac{\alpha}{2\pi} \int_0^{2\pi}d\theta' \, P(\boldsymbol{r},\theta',t)
        \label{eq:FPE}
    \end{split}
\end{equation}
where for an ABP one sets $\alpha = 0$ in the above expression, and respectively for an RTP one sets $D_r = 0$.  
To find $a_0(\boldsymbol{r},t)$, it is sufficient to find the marginal of $P(\boldsymbol{r},\theta,t)$, \textit{i.e.} $P(\boldsymbol{r},\theta,t)$ integrated over all values of $\theta$. To get rid of $\theta$ in this FPE, we closely follow the method of Dinelli et al.~\cite{dinelli_fluctuating_2024} for its simplicity when working with 2D active particles. Note, however, that we use a different closure for our purpose.
The angular dependences are separated from the radial ones via a Fourier series
\begin{equation}
\label{eq:decomposition harmonics}
    P(\boldsymbol{r},\theta,t)=\frac{1}{2\pi}\sum_{n=-\infty}^{+\infty}e^{in\theta}a_n(\boldsymbol{r},t)
\end{equation}
introducing the different coefficients $a_n(\bmr,t)$ for $n \in \mathbf{Z}$. Introducing the scalar product $\langle f,g\rangle = \int_0^{2\pi}d\theta \;f^*(\theta)g(\theta)$, we note that each component may be obtained as an integral over the mode $n$, as
\begin{align*}
    \langle e^{in\theta},P\rangle &= \int_0^{2\pi}d\theta \;e^{-in\theta}\frac{1}{2\pi}\sum_{k=-\infty}^{+\infty}e^{ik\theta}a_k(\boldsymbol{r},t) = a_n.
\end{align*}
We now aim to reexpress the FPE Eq.~\eqref{eq:FPE} as a function of the $a_n(\bmr,t)$ so as to obtain $a_0(\bmr,t)$. To do so, a few more relationships are worth deriving:
%We can then follow~\cite{dinelli_fluctuating_2024}, and with the scalar product $\langle f,g\rangle = \int_0^{2\pi}d\theta \;f^*(\theta)g(\theta)$ and the vectors $\hat{i}_\pm=\begin{pmatrix} 1 \\ \pm i\end{pmatrix}$, we show the following relationships :
\begin{align*}
    \langle e^{in\theta},\partial^2_\theta P\rangle &= \int_0^{2\pi}d\theta \;e^{-in\theta}\frac{1}{2\pi}\sum_{k=-\infty}^{+\infty}-k^2e^{ik\theta}a_k(\boldsymbol{r},t) \\&=-n^2a_n \\
    \langle e^{in\theta},\boldsymbol{u}(\theta)P\rangle &= \int_0^{2\pi}d\theta \;e^{-in\theta}\frac{1}{2\pi}\sum_{k=-\infty}^{+\infty}\begin{pmatrix} \cos{\theta} \\ \sin{\theta} \end{pmatrix}e^{ik\theta}a_k(\boldsymbol{r},t) \\ &=\frac{1}{2}\left[ \hat{i}_+a_{n+1}+\hat{i}_- a_{n-1}\right]
\end{align*}
where we also introduced the vectors $\hat{i}_\pm=\begin{pmatrix} 1 \\ \pm i\end{pmatrix}$. In addition we note that 
\begin{align*}
    \int_0^{2\pi} d\theta \,  P(\boldsymbol{r},\theta,t) &=  \int_0^{2\pi} d\theta \,  \frac{1}{2\pi}\sum_{n=-\infty}^{+\infty}e^{in\theta}a_n(\boldsymbol{r},t) \\ &=  \sum_{n=-\infty}^{+\infty} \, a_n \int_0^{2\pi} d\theta \frac{e^{in\theta}}{2\pi} \\
    &= \sum_{n=-\infty}^{+\infty}  \, a_n \, \delta_{n,0} =  a_0.
\end{align*}
We can then straightforwardly compute $\langle e^{in\theta},\partial_tP\rangle$ and obtain a series of equations for the evolution of all the $a_n$ coefficients,
\begin{equation}
\begin{split}
    \partial_ta_n=-\frac{v}{2}\nabla_r\cdot \left[ \hat{i}_+a_{n+1}+\hat{i}_- a_{n-1}\right] + D_t\Delta a_n  \\ - D_r n^2 a_n - \alpha(1-\delta_{n,0})a_n.\end{split}
    \label{eq:coefficients}
\end{equation}
In the following it will be useful to compare different levels of truncation of these series and so we explicitly write out the first three equations of this series. Note, that we also add a term as $\delta(t)\delta(\boldsymbol{r})$ in $a_0$ to account for initial conditions starting at $t=0$ and $\boldsymbol{r} = \boldsymbol{0}$ : 
\begin{equation}
\begin{split}
    \partial_t a_0(\boldsymbol{r}, t)
    = -\frac{v}{2} \nabla_r \cdot \left[ \hat{i}_+ a_{1} + \hat{i}_- a_{-1} \right] + D_t \Delta a_0 \\
    \quad + \delta(t) \delta(\boldsymbol{r}),
\end{split} \label{equ : a+-1}
\end{equation}
\begin{equation}
\begin{split}
    \partial_t a_{\pm1}(\boldsymbol{r}, t)
    = -\frac{v}{2} \nabla_r \cdot \left[ \hat{i}_\pm a_{\pm2} + \hat{i}_\mp a_{0} \right] + D_t \Delta a_{\pm1} \\
    \quad - (D_r + \alpha) a_{\pm1},
\end{split} \label{equ : a+-1}
\end{equation}
\begin{equation}
\begin{split}
    \partial_t a_{\pm2}(\boldsymbol{r}, t)
    = -\frac{v}{2} \nabla_r \cdot \left[ \hat{i}_\pm a_{\pm3} + \hat{i}_\mp a_{\pm1} \right] + D_t \Delta a_{\pm2} \\
    \quad - (4D_r + \alpha) a_{\pm2}.
\end{split} \label{eq:coefficients 2}
\end{equation}
% \begin{align}
%      \partial_ta_0(\boldsymbol{r},t)&=-\frac{v}{2}\nabla_r\cdot \left[ \hat{i}_+a_{1}+\hat{i}_- a_{-1}\right] + D_t\Delta a_0 \\& \phantom{-\frac{v}{2}\nabla_r\cdot \left[ \hat{i}_\pm a_{\pm2}+\hat{i}_\mp a_{0}\right]}+\delta(t)\delta(\boldsymbol{r}) \notag\\
%     \partial_ta_{\pm1}(\boldsymbol{r},t)&=-\frac{v}{2}\nabla_r\cdot \left[ \hat{i}_\pm a_{\pm2}+\hat{i}_\mp a_{0}\right] + D_t\Delta a_{\pm1} \label{equ : a+-1} \\& \phantom{-\frac{v}{2}\nabla_r\cdot \left[ \hat{i}_\pm a_{\pm2}+\hat{i}_\mp a_{0}\right]} - (D_r+\alpha) a_{\pm1} \notag  \\
%     \partial_ta_{\pm2}(\boldsymbol{r},t)&=-\frac{v}{2}\nabla_r\cdot \left[\hat{i}_\pm a_{\pm3} + \hat{i}_\mp a_{\pm1}\right] + D_t\Delta a_{\pm2}  \label{eq:coefficients 2} \\& \phantom{-\frac{v}{2}\nabla_r\cdot \left[ \hat{i}_\pm a_{\pm2}+\hat{i}_\mp a_{0}\right]} - (4D_r+\alpha) a_{\pm2} \notag
% \end{align}

The approach on the infinite series of $a_n$ can be related to other techniques consisting in obtaining evolution equations for the fields describing the density $\rho$, the polar $\boldsymbol{p}(\boldsymbol{r},t)$, and nematic orders of the system $  \underline{\underline{q}}(\boldsymbol{r},t)$ \cite{ahmadi_hydrodynamics_2006, gautry_closures_2025, marchetti2013hydrodynamics, martin_statistical_2021, burekovic_active_2026, spera_nematic_2024}. 
The advantage of working at the level of the $a_n$ is that the $n$ coefficients are directly related to one another through a recurrence relation while the link with higher orders when working with these fields is not obvious. Although the physical meaning of the $a_n$ coefficients is less straightforward than the meaning of these fields, it is possible to connect these approaches, via the relations
\begin{align*}
    \rho(\boldsymbol{r},t) &= \int P(\boldsymbol{r},\boldsymbol{u},t) \; d\boldsymbol{u} = a_0,\\
    \boldsymbol{p}(\boldsymbol{r},t) &= \int \boldsymbol{u}P(\boldsymbol{r},\boldsymbol{u},t) \; d\boldsymbol{u} = \begin{pmatrix}
        a_1+a_{-1} \\ i(a_1 - a_{-1}) 
    \end{pmatrix}, \\
    \underline{\underline{q}}(\boldsymbol{r},t) &= \int \left( \boldsymbol{u}  \boldsymbol{u}-\frac{\mathbb{I}}{2} \right) \, P(\boldsymbol{r},\boldsymbol{u},t) \; d\boldsymbol{u} \\ &= \begin{pmatrix}
        \frac{1}{2}(a_2 + a_{-2}) &i(a_2 + a_{-2} ) \\ i(a_2 + a_{-2} ) & -\frac{1}{2}(a_2 + a_{-2})
    \end{pmatrix},
\end{align*}
shown in Appendix.~\ref{appendix:equivalence}. These relations show in particular that $a_0(\bmr,t)$ represents the probability density of finding particles in $\bmr$, $\rho(\bmr,t)$. 

We note that although their notation in the literature are very similar, the density $\rho(\boldsymbol{r},t)$ is a different object than the microscopic density which characterizes the ensemble of individual particle positions. One is a probability density that is a macroscopic, smooth object corresponding to a distribution of positions while the other is a sum of delta peaks corresponding to a microscopic function that needs to be handled with more care. We will stick with the $a_0$ notation instead of $\rho(\boldsymbol{r},t)$ as a reminder that it is a probability obtained through an expansion.

To make progress, it will be useful in the following to use the time and space Fourier transform $$\widetilde{a}_{n}(\kk,\omega) = \iint \dd \bmr \, \dd t \, a_n(\bmr,t) e^{-i\omega t} e^{-i\kk \cdot \bmr}.$$ 

\paragraph{Closure procedure and infinite continued fraction.} 
Eq.~\eqref{eq:coefficients} describes an infinite hierarchy of equations, with each $a_n$ coefficient depending on $a_{n \pm 1} $. A truncation will therefore be necessary to obtain expressions for $a_0$. A standard procedure~\cite{ahmadi_hydrodynamics_2006,marchetti2013hydrodynamics, dinelli_fluctuating_2024, solon2015active, martin_statistical_2021, gautry_closures_2025} consists in cutting higher orders, $n > N$, corresponding to either the polar, the nematic, or higher angular dependences. Truncating at order $N$ implies taking all components $\{a_{\pm n}\}_{n \geq N+1} =0$. With a truncation at order $N$, using Eq.~\eqref{eq:coefficients}, the Fourier components $\widetilde{a}_{\pm N}$ can then be expressed as
\begin{align*}
    i\omega\widetilde{a}_{\pm N}(\boldsymbol{k}, \omega) = -\frac{v}{2} i\boldsymbol{k} \cdot \hat{i}_{\mp} \widetilde{a}_{\pm (N-1)} - D_tk^2\widetilde{a}_{\pm N} \\ - N^2 D_r \widetilde{a}_{\pm N} -\alpha\widetilde{a}_{\pm N},
\end{align*}
This expression may be recast into
\begin{align*}
    \widetilde{a}_{\pm N}(\boldsymbol{k}, \omega) = \frac{-\frac{v}{2} i\boldsymbol{k} \cdot \hat{i}_{\mp} \widetilde{a}_{\pm (N-1)}}{b_N},
\end{align*}
introducing the notation
\begin{align}
    b_n &= i\omega + D_tk^2 + n^2D_r + \alpha(1 - \delta_{n,0}), \,\, \forall n \in \mathbf{N}.
    \label{eq:bn}
    \end{align}
Terms of order $N-1$ are then found iteratively
\begin{equation}
    \begin{split}
i\omega&\widetilde{a}_{\pm (N-1)}= -\frac{v}{2} i\boldsymbol{k} \cdot \hat{i}_{\mp} \widetilde{a}_{\pm (N-2)} \\ & -\left(\frac{\frac{v^2k^2}{4}}{b_N} +D_tk^2 + (N-1)^2 D_r  +\alpha \right)\widetilde{a}_{\pm (N-1)}
\end{split}
\end{equation}
which can be recast into
\begin{equation*}
    \begin{split}
 \widetilde{a}_{\pm (N-1)} = \frac{- \frac{v}{2} i\boldsymbol{k} \cdot \hat{i}_{\mp} \widetilde{a}_{\pm (N-2)}}{b_{N-1} + \frac{c_N}{b_N}}
    \end{split}
\end{equation*}
introducing the notation
\begin{align}
    c_n & = \frac{v^2k^2}{4}, \,\, \forall n \in \mathbf{N}.
    \label{eq:cn}
\end{align}
Continuing this scheme, it is clear we can express all $\widetilde{a}_{n\leq N}$ as a function of $c$ and $b_n$. We thus introduce the notation for continued fraction
\begin{align*}
     \K_{n=1}^N{\frac{c_n}{b_n}} =  \cfrac{c_1}{b_1 +\cfrac{c_2}{b_2 + \cfrac{c_3}{b_3 + \ddots}}}
\end{align*}
such that for $\widetilde{a}_{\pm1}$,
\begin{equation*}
    \widetilde{a}_{\pm1} = \frac{-\frac{v}{2}i\boldsymbol{k}\cdot \widetilde{a}_0}{b_1+\K_{n=2}^N {\frac{c_n}{b_n}}}.
\end{equation*}
Finally an expression for $\widetilde{a}_0$ correct to order $N$ in the truncation is 
\begin{equation}
\label{eq:a_0 continued fraction full}
    \widetilde{a}_0^{-1} = b_0 + 2\K_{n=1}^N{\frac{c_n}{b_n}}.
\end{equation}
This compact expression is one of the main results of our work. Notice that when we go to the limit of a passive tracer, with $v = 0$, we recover $\widetilde{a}_0 (\kk,\omega) = 1/(i\omega + D_t k^2)$ which is the standard expression for a purely diffusive particle~\cite{hansen2013theory}. 

In the case of RTPs, Eq.~\eqref{eq:a_0 continued fraction full} can be further simplified. Since neither the denominators nor numerator depend on $n$ in this case, as given by Eqs.~\eqref{eq:bn} and \eqref{eq:cn}, $b_n \equiv b$ and $c_n \equiv c$, the continued fraction converges to an analytic expression when taking the limit of $N\rightarrow \infty$, which is
\begin{align*}
    \widetilde{a}_0^{-1} &= b_0 + \K_{n=1}^\infty \frac{c}{b} = b_0 + \frac{4c}{b+\sqrt{b^2+4c}}. 
\end{align*}
Using the expressions for $b$ and $c$ as provided in Eqs.~\eqref{eq:bn} and \eqref{eq:cn} yields 
\begin{equation}
   \widetilde{a}_0(\kk,\omega)  = \frac{1}{\sqrt{(i\omega + D_tk^2 +\alpha)^2+v^2k^2}- \alpha}
   \label{eq:a_0 RTPs}
\end{equation}
which is another main result of our work. 
This expression is analytically exact as all the terms are accounted for and the hierarchy can be closed without the need for a truncation (since we took $N \rightarrow \infty$). This is, to our knowledge the first exact result obtained for the probability distribution for non-interacting RTPs in 2D from a perturbative series on the orientational orders. Our result in Eq.~\eqref{eq:a_0 RTPs} is similar to Eq.~(20) of Ref.~\cite{martens_probability_2012}. In comparison, Ref.~\cite{martens_probability_2012} does not contain translational diffusion, which can actually be added in the entire derivation by replacing $i \omega \rightarrow i \omega + D_t k^2$. Ref.~\cite{martens_probability_2012} relies on a perturbative series on the number of tumbling events~\cite{hauge1970exact}, rather than on the distribution of orientational orders. As such, it is not easily transferable to other choices of reorientation dynamics.

For ABPs, one has no choice but to choose an order at which to close the series. Taking a truncation at order $N$ essentially corresponds to choosing the value for which $\frac{c}{b_{N+1}}$ will be negligible compared to $b_N$. Therefore, truncation will be more accurate when $c \ll b_n^2$. A competition between the values of the speed $v$ in $c$ and of the diffusion coefficients in $b_n$ at order $k^2$ can therefore be identified, and $\frac{c}{b_n^2}$ will be smaller at smaller $\text{Pe}$. The truncation will thus also converge quicker for ABPs, due to the factor $n^2$ appearing in front of $D_r$ in the definition of $b_n$ in Eq.~\eqref{eq:bn}. This contribution comes from the diffusive reorientation in space for ABPs and is absent in RTPs for which reorientation dynamics are sudden. As such, the truncation consisting in neglecting higher orientational moments will be more accurate for systems where the orientation avoids sudden changes and evolves smoothly.
%In order to close it we rely on approximations. We will not take into account the fast field approximation~\cite{solon2015active,martin_statistical_2021,dinelli_fluctuating_2024} \sophie{which one? Not defined earlier, be precise, cite literature} consisting in taking $\{\partial_ta_n \}_{n \geq 1} =0$ by relying on the fact that all harmonics above $a_0$ have exponential decay, as we are interested in a full time description of the ISF of our systems for which the low times impact of these coefficients is essential. We can however take an approximation describing the systems at large enough scales and truncate higher harmonics, as the divergence operator makes each harmonics impact the density at smaller scale than the previous ones. \sophie{not sure I understand how this saves the short times. also you should relate to Gautry here. Maybe these differences are worth mentioning in the intro to point out what differences we make between our work and previous work}

\begin{widetext}
For ABPs we can for instance stop at polar order, taking $N =1$ such that $\{ a_n \}_{n\geq 2} = 0$, and obtain
\begin{equation}
\label{eq:a0 order1}
    \widetilde{a}_{0,1} (\boldsymbol{k},\omega) = \frac{i\omega + D_tk^2+D_r+\alpha}{(i\omega+D_tk^2)(i\omega+D_tk^2+D_r+\alpha)+\frac{v^2k^2}{2}},
\end{equation}
where we used the notation $ \widetilde{a}_{0,N}$ to designate $\widetilde{a}_{0}$ approximated with a truncation at order $N$. 
Stopping at nematic order, taking $N= 2$ and $\{ a_n \}_{n\geq 3} = 0$, yields
\begin{equation}
\label{eq:a0 order2}
    \widetilde{a}_{0,2}(\boldsymbol{k},\omega) = \frac{2(i\omega +D_tk^2 +4D_r +\alpha)(i\omega +D_tk^2 +D_r +\alpha)+\frac{v^2k^2}{2}}{(i\omega +D_tk^2)[2(i\omega +D_tk^2 +4D_r +\alpha)(i\omega +D_tk^2 +D_r +\alpha)+\frac{v^2k^2}{2}]+v^2k^2(i\omega+D_tk^2+4D_r+\alpha)}.
\end{equation}
\end{widetext}

\subsection{Intermediate Scattering Functions (ISF)}
\label{sec:ISF}

\begin{figure*}[t!]
    \centering
    \includegraphics[width=0.99\linewidth]{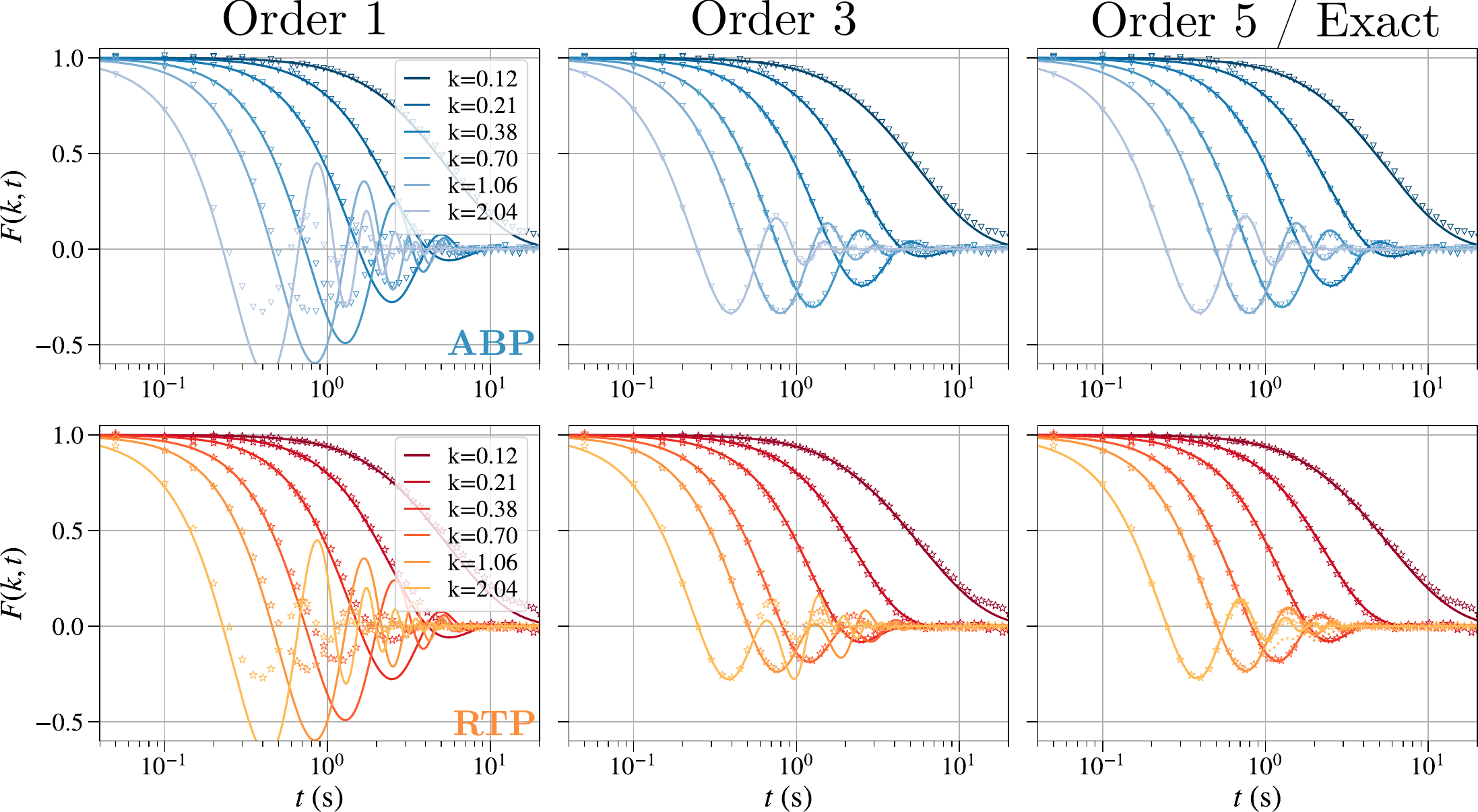}
    \caption{\textbf{Intermediate Scattering Functions} for Active Brownian Particles (top), and Run-and-Tumble particles (bottom) obtained at different orders in the truncation. k size are taken to be the same for all three plots and go from small length scales (in light colors) to large ones (in dark colors). Dynamic parameters are $v = 5~\unit{\mu m.s^{-1}} $, $D_t = 0.1~\unit{\mu m^2.s^{-1}} $, and $D_r = 1~\unit{s^{-1}}$. For order 5/Exact, in the top ABP graph the full line corresponds to an order $N = 5$ truncation, while in the bottom RTP plot, the full line is the exact result from the inverse FFT of Eq.~\eqref{eq:a_0 RTPs} and the dotted line from an order $N = 5$ truncation.}
    \label{fig:ISF_ABP_RT}
\end{figure*}

% For the case of ABPs and RTPs we will then use this preliminary work on the obtention on equations for the harmonics in Fourier space to approximate the ISF. This method has been proposed recently by Gautry et al. [CITE PIERRE AND TIMOTHE PREPRINT] who use it to study anisotropic ABPs and show the high precision of this approximation within a reasonable Péclet number range.
% \par
% Looking back on Eq.~\eqref{eq:ISF nonin} we can seen that we can express $F(\boldsymbol{k},t)$ in terms of $a_0$ when the particles are non interacting. Furthermore we can see that the part depending on initial conditions $\boldsymbol{r'}$ in the ISF is expressed as $e^{i\boldsymbol{k}\cdot\boldsymbol{r'}}$ and appears in $a_0(\boldsymbol{k},t)$ as a prefactor $e^{-i\boldsymbol{k}\cdot\boldsymbol{r'}}$. These two contributions in $\boldsymbol{k}$ space therefore cancel out and the ISF is independent of initial conditions as one can expect. It can then simply be calculated with the assumption of initial conditions $\boldsymbol{r'} = 0$ without loss of generality and we can then see that we have 
% \begin{equation*}
%     F(\boldsymbol{k},t) = \int d\boldsymbol{r} e^{-i\boldsymbol{k}\cdot \boldsymbol{r}}a_0(\boldsymbol{r},t)
% \end{equation*}
% We can then easily calculate $F(\boldsymbol{k},t) $ from our expressions for $\widetilde{a}_0(\boldsymbol{k},\omega)$ by simply inverting the Fourier transform on time.

\paragraph{From frequency domain to real time.}
Using Eq.~\eqref{eq:ISF def}, we can relate $\tilde{a}_0(\kk, \omega)$ to the self ISF doing an inverse Fourier transform in time $F(\kk,t) = \int\frac{\dd \omega}{2\pi}  \, e^{i\omega t} \tilde{a}_0(\kk, \omega) $.
At order $N=1$ in the expansion, or polar order, we can simply take the inverse Fourier transform of Eq.~\eqref{eq:a0 order1}. %in the backward this gives us the following expression,
%\begin{equation}
%    F_1(\boldsymbol{k},t) = \int_{-\infty}^{\infty} \frac{\dd\omega}{2\pi}\frac{e^{i\omega t}(i\omega + D_tk^2+D_r+\alpha)}{(i\omega+D_tk^2)(i\omega+D_tk^2+D_r+\alpha)+\frac{v^2k^2} {2}} 
%\end{equation}
This integral can be solved using the Residue Theorem, with two poles canceling the denominator,
\begin{equation*}
    \omega_\pm= i(D_tk^2+D_r)\pm \sqrt{\frac{k^2v^2}{2}-\frac{D_r^2}{4}}
\end{equation*}
which yields
\begin{equation}\begin{split}
    F_1&(\boldsymbol{k},t) = e^{-(D_tk^2+\frac{D_r^{\alpha}}{2})t}\Bigg[  \mathrm{\cos}\left(\frac{t}{2}\sqrt{2k^2v^2-(D_r^{\alpha})^2}\right)\\& + \frac{D_r^{\alpha}}{2\sqrt{2k^2v^2-(D_r^{\alpha})^2}} \mathrm{\sin}\left(\frac{t}{2}\sqrt{2k^2v^2-(D_r^{\alpha})^2}\right)\Bigg] 
\end{split}\end{equation}
where we abbreviated $D_r^{\alpha} = D_r + \alpha$. The fact that we can factor out $D_r$ and $\alpha$ in the same way means that at this order of the truncation, $D_r$ and $\alpha$ play the same role. This is also expected from looking at $a_{\pm 1}$ terms in Eq.~\ref{equ : a+-1} where $D_r$ and $\alpha$ play the same role. The polar order truncation thus smears out differences between reorientation modes.

The same strategy can be used to obtain an ISF at nematic order. The denominator is of order 3 in this case and we find 3 poles $(\omega_1,\omega_2,\omega_3)$ in the superior part of the complex plane (see Appendix.~\ref{appendix:math details} for details). Writing $h(\omega) = (i\omega + D_tk^2+D_r+\alpha)(i\omega +D_tk^2 +4D_r+ \alpha) + \frac{v^2k^2}{4}$, we get
\begin{align}
    F_2(\boldsymbol{k},t) = -\sum_{i=1}^3 \frac{h(\omega_i)e^{i \omega_i t}}{\prod_{j \neq i} (\omega_i - \omega_j)}.
    \label{eq:f2kt}
\end{align}

For higher orders, it gets increasingly harder to obtain analytical results. 
Eq.~\eqref{eq:a0 order1} and \eqref{eq:a0 order2} highlight that the denominator of $\widetilde{a}_0(\kk,\omega)$ when the truncation is stopped at order $N$ will be of order $N+1$ in $\omega$. Therefore, analytic solutions for $F(\boldsymbol{k},t)$ can only be found up to order $N=3$ in the truncation as only polynomials of up to order 4 admit general solutions for their roots. For higher orders, one thus has to invert the Fourier transform numerically, with fast Fourier transform (FFT) algorithms. We do not solve at order 3 analytically in this paper for the sake of brevity, and all the figures with order $N \geq 3$ are obtained from perfoming that numerical integration. 

In the case of RTP, where we have an exact expression for $\widetilde{a}_0$ given via Eq.~\eqref{eq:a_0 RTPs}, we also have to perform a numerical inverse Fourier transform. 
%we have the integral :
%\begin{equation}
%     F(\boldsymbol{k},t) = \int_{-\infty}^{\infty} \frac{\dd \omega}{2\pi} \,  \frac{e^{i\omega t}}{- \alpha+\sqrt{(i\omega + D_tk^2 +\alpha)^2+v^2k^2}}
%\end{equation}
%It doesn't admit an analytical solution \tristan{DOES IT NOT I'M LESS SURE ABOUT THIS ONE} but is readily computed numerically. 
Numerically obtaining $F(\kk,t)$ from $\widetilde{a}_0$ with FFTs is a fast and robust process, compared to alternatives which rely on numerically solving a system of equations ~\cite{zhao_quantitative_2024}. Inspiring from Ref.~\cite{martens_probability_2012}, we can reexpress Eq.~\ref{eq:a_0 RTPs} as 
\begin{equation*}
    \widetilde{a}_0 = \frac{P_0}{1-\alpha P_0} = \sum_{n=0}^{\infty} \alpha^n P_0^{n+1}
\end{equation*}
with $P_0 = \sqrt{(i\omega + D_tk^2 + \alpha)^2 + v^2k^2}^{-1}$. The inverse Fourier transform can then be carried out within the sum and we find a rather straightforward expression, in the case of RTP only, as
\begin{equation*}
    F(\bm{k},t) = e^{-\alpha t - D_tk^2t} \sum_{n=0}^{\infty} \frac{\sqrt{\pi}}{2^{\frac{n}{2}} \Gamma\left(\frac{n+1}{2}\right)} \left( \frac{\alpha^2 t}{k v} \right)^{\frac{n}{2}} J_{\frac{n}{2}}(k v t)
\end{equation*}
where the $J_{n/2}$ are Bessel functions of the first kind.
In practice, this direct expression in real time could be computationally advantageous. However, it requires to conduct the sum at high values of $n$ where commonly used algorithms to calculate Bessel functions are inaccurate~\cite{bessel}.
%\tristan{an inverse Laplace is essentially an inverse FFT, not sure it can be done as fast as the FFT. But I think we have to be careful with the criticism} \tristan{This is criticism that Christina et al do themselves in \cite{zhao_quantitative_2024}, in fact they numerically solve like the 8 integrals of the Renewal technique instead of doing an Inverse Laplace transform even though there is an analytical exact result in Laplace space because it is faster. We can also cite \cite{craig_practical_1994} which is a nice computational article explaining the struggles with Laplace transforms}

%\begin{figure}[h!]
%    \centering
%    \includegraphics[width=0.99\linewidth]{fig2_PRE_vertical.pdf}
%    \caption{\textbf{Intermediate Scattering Functions} for (a) Active Brownian Particles, (b) Run-and-Tumble particles, and (c) Active Ornstein-Uhlenbeck particles. k size are taken to be the same for all three plots and go from small length scales (in light colors) to large ones (in dark colors). Dynamic parameters are $v = 5~\unit{\mu m.s^{-1}} $, $D_t = 0.1~\unit{\mu m^2.s^{-1}} $, and $D_r = 1~\unit{s^{-1}}$ }
%    \label{fig:ISF}
%\end{figure}

\paragraph{Convergence of the obtained expressions.}

We now compare our analytical predictions for $F(\kk,t)$ with measured ISFs from simulated data, both for ABP and RTP (Fig.~\ref{fig:ISF_ABP_RT}). We especially compare the results at different orders $N$. 
%The order 2 ISFs obtained at different orders for ABPs and RTPs can be seen on Fig.~\ref{fig:ISF_ABP_RT}. 
Since our particles do not interact, we are effectively looking at the self ISF which satisfies $F(\boldsymbol{k},0) = 1$ and decays towards zero in time. For both RTP and ABP, we find this decay is accompanied with significant oscillations -- a fact that has been identified though seldom explained in several works prior ~\cite{kurzthaler2018probing,zhao_quantitative_2024}. 

To understand the emergence of these oscillations, a key lies in the convergence of our predictions at increasing truncation order $N$. 
Indeed, the main improvement between simulation data and theory at higher orders is in the description of the oscillations. At order 1 the theory overestimates their amplitude and they decay slower than they should, as higher angular harmonics are neglected. A particle's rotation is therefore not accurately accounted for and the ISF behaves as if particles are assuming a longer persistence time in one direction then what is actually the case. Order 3 already describes ABP motion quite accurately -- at least for the set of investigated parameters -- while there are still some minor differences for RTPs. This is an expected consequence as we know that the infinite continued fraction converges slower in the case of RTPs. Eventually, the exact result for RTPs is fully converged and describes the motion perfectly at all length and time scales. 
This convergence of how well we describe oscillations hints to the fact that these oscillations must originate from persistent motion in one direction. 

\begin{figure}[h]
    \centering
    \includegraphics[width=0.85\linewidth]{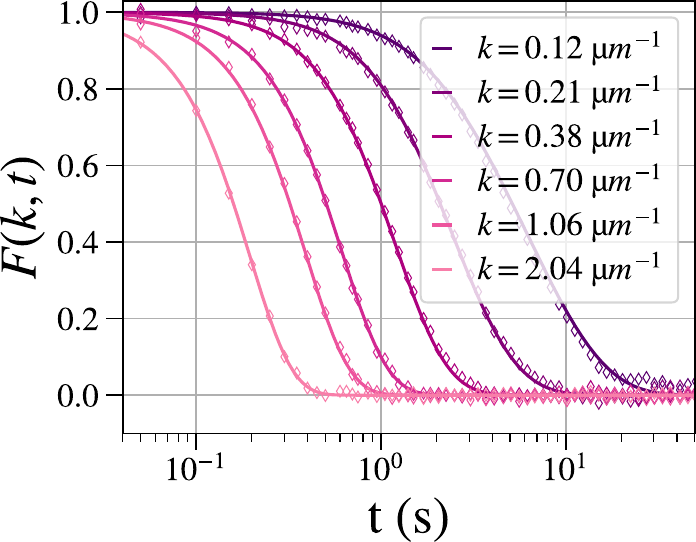}
    \caption{\textbf{Intermediate Scattering Function} of Active Ornstein-Uhlenbeck Particles. The dynamic parameters used are equivalent to the ones in Fig~\ref{fig:ISF_ABP_RT}.}
    \label{fig:ISF_AOUP}
\end{figure}

Overall, oscillations do not appear for small $k$ and are damped at long times when the dynamics can be described effectively as a diffusive process. For the AOUP model (Fig.~\ref{fig:ISF_AOUP}), the ISF is a simple decreasing exponential exhibiting no oscillations. The oscillations of the ABP and RTP models are thus a marker of non gaussianity. Similar oscillations can be observed in scattering measurements of particles moving with a mean drift such as sedimenting particles \cite{levitz_probing_2025}, and is a hallmark of the ISF found from the propagator of free running particles \cite{li_detecting_2025}. The oscillations must thus come from the persistent motion in one direction, and naturally, relax differently according to the different reorientation mechanisms.

\section{Countoscope for self-propelled particles}
\label{sec:countoscope}

\subsection{Number Fluctuations}

The obtain the number correlation function $C_N(t)$ and the NMSD $\langle (N(t) - N(0))^2 \rangle$, we can use our predictions for $F(\kk,t)$ and simply input them in Eq.~\eqref{eq:Cnt}, meaning taking a weighted integral over $\kk$. Apart from the AOUPs, for which we have seen an explicit analytical solution exists to this integral (see Eq.~\eqref{eq:Cnt Gaussian}), for ABPs and RTPs, one has to perform a numerical integration. Using the numerical scheme detailed in Appendix.~\ref{appendix:computational methods}, this can be done quickly at little computational expense.

\begin{figure*}
    \centering
    \includegraphics[width = 0.99\linewidth]{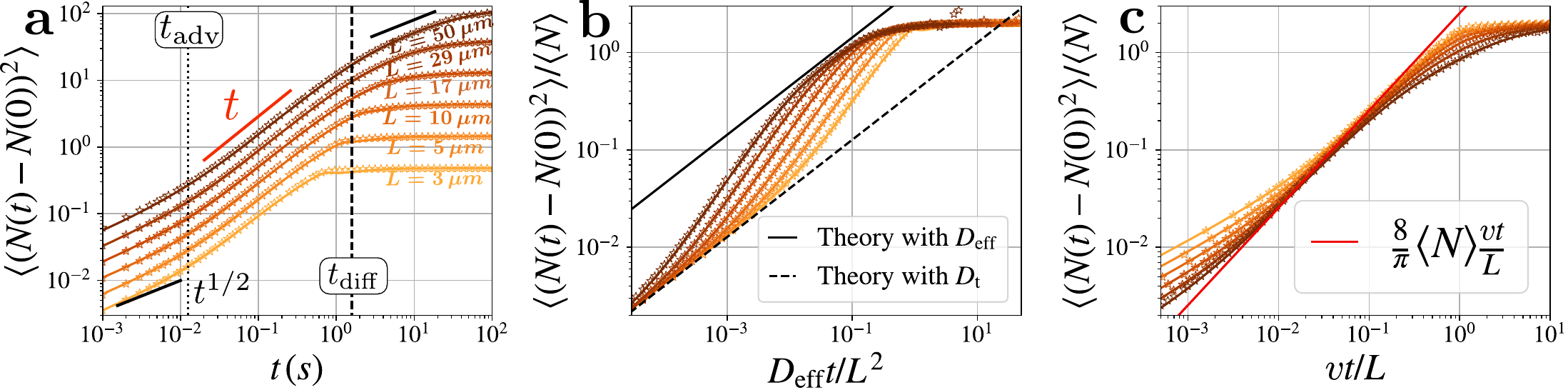}
    \caption{\textbf{NMSD behavior for Run-and-Tumble particles.} (a) NMSD behavior with respect to lag time for increasing box sizes going from yellow to brown; (b) same as (a) but where time is rescaled by a typical diffusion timescale $L^2/D_{\rm eff}$ and the NMSD by $\langle N \rangle$, and (c) same as (b) but where time is rescaled by a typical advection time $L/v$. In all of the plots, stars correspond to numerical simulations and lines to the theory obtained from integrating $F_2(k,t)$ given by Eq.~\eqref{eq:f2kt} inputted in Eq.~\eqref{eq:Cnt}. Physical parameters are the same as in Fig.~\ref{fig:ISF_ABP_RT}. Here we have $t_{\rm adv} = \pi \frac{D_t}{v^2}$ and $t_{\rm diff} = \pi \frac{D_{\rm eff}}{v^2}$ the crossover times between regimes for the NMSD.}
    \label{fig:All NMSD}
\end{figure*}

As a canonical illustration of the behavior of the number fluctuations, we present in Fig.~\ref{fig:All NMSD}-a, the NMSD curves for RTPs using the analytical expressions at order $N = 2$, for different box sizes (lines). We compare these with results from numerical simulations (stars) and find excellent agreement between the two. The NMSD exhibits several peculiar features. First, the number fluctuations increase in time, before reaching a plateau. Starting from an initial number configuration, the longer one observes particles inside a box the more likely one is to sample number configurations further away from the initial configuration; hence the number fluctuations increase. The plateau corresponds to the time when the memory of the initial configuration is lost, essentially when correlations in the number fluctuations have decayed sufficiently. The value of the plateau is given by the variance of the particle number $\lim_{t \to \infty} \langle \Delta N(t) ^2 \rangle  = 2 (\langle N^2 \rangle - \langle N  \rangle ^2)$, as one can see from Eq.~\eqref{eq:counting}. In a non-interacting suspension, $\langle N^2 \rangle - \langle N  \rangle ^2 = \langle N \rangle$. The plateau behavior is confirmed as all NMSDs collapse at long times when they are rescaled by $\langle N \rangle$ (Fig.~\ref{fig:All NMSD}-b and c). 
The plateau is reached faster for smaller boxes where correlations are smaller, as initial particles are less likely to stay in the box or come back to the box. 

The NMSD exhibits three different regimes in time, similarly to the MSD of RTP \cite{howse_self-motile_2007}. First, growing as $t^{1/2}$, then increasing faster as $t$ at intermediate times and slower as $t^{1/2}$ at long times. Different rescalings can reveal this phenomena more accurately. First, we rescale time by $L^2/D_{\rm eff}$, the time it takes to diffuse across a box (c), and curves collapse at short and long times. Second, we rescale time by $v/L$ , the time it takes for a particle to be advected across the box (b) and find excellent collapse of NMSDs at intermediate time. These regimes thus correspond to the two diffusive regimes at short and long time and the advective regime at intermediate time known for these self-propelled models. Notice that all regimes in NMSD space grow in time typically with a power that is the half of the power seen in the MSD. Whereas in a MSD diffusive behaviour scales linearly with $t$ and advective as $t^2$, diffusive regimes in the NMSD increase as $t^{1/2}$ and the advective one as $t$. Overall, our theoretical predictions for the NMSD recover all the dynamical aspect of self-propelled particles, demonstrating their potential to probe and investigate dynamical behavior of self-propelled suspensions. %\tristan{Not sure at all how to write this part and what to write in it this shit is hard man, there's probably way more to say but also maybe should be said elsewhere ?}

\subsection{Limiting Regimes}

Further physical and analytical insights may be drawn by investigating the different limiting regimes of the NMSD.

\paragraph{Gaussian case.} For Gaussian models this can easily be done, as the NMSD can be expressed directly in terms of the MSD, and Taylor expansions may readily be obtained at different timescales. At short time scales, Eq.~\eqref{eq:Cnt Gaussian} becomes
\begin{align}
    &C_N(t) \simeq \langle N \rangle \left( 1- \frac{2}{\sqrt{\pi}}\sqrt{\frac{\langle \Delta r^2 (t) \rangle}{L^2}} \right), \\
    & \langle \Delta N^2(t) \rangle \simeq \langle N \rangle \frac{4}{\sqrt{\pi}}\sqrt{\frac{\langle \Delta r^2 (t) \rangle}{L^2}}. 
\end{align}
One immediately sees that the exponents in time of the MSD are thus halved in the NMSD space. 
Plugging in the limiting expressions of the MSD (advective or diffusive) thus yields limiting expressions for the NMSD. For the case of AOUPs, diffusive regimes with diffusive coefficient $D$ (where $D=D_t$ at short timescales or $D = D_{\rm eff}$ at longer timescales) can thus be expressed as
\begin{equation}
\label{eq: diffusive regime}
    \langle \Delta N^2(t) \rangle \simeq \frac{8}{\sqrt{\pi}} \langle N \rangle \sqrt{\frac{Dt}{L^2}}
\end{equation}
and the advective regime as
\begin{equation}
\label{eq:advective regime AOUP}
    \langle \Delta N^2(t) \rangle \simeq \frac{4}{\sqrt{\pi}} \langle N \rangle \frac{vt}{L}.
\end{equation}

\paragraph{Diffusive limits in non Gaussian cases.} For a non gaussian model, limiting regimes can also be obtained but require more subtle investigations.  
For ABPs and RTPs, the first limiting regime can be obtained by neglecting the effect of activity, which is equivalent to truncating at order $N=0$ in Eq.~\eqref{eq:a_0 continued fraction full}. This corresponds to a passive diffusive tracer with diffusion coefficient $D_t$. One then obtains a prediction for $C_N(t)$ given by Eq.~\eqref{eq:Cnt Gaussian} where $\langle \Delta r^2 (t)\rangle = 4 D_t t$; and hence we obtain at short times the same result as for AOUPs, given by Eq.~\eqref{eq: diffusive regime} with $D = D_t$.

Long timescales and large length scales can be described through an alternative construction of $a_0$. Following the ``fast field'' truncation procedure used in~\cite{dinelli_fluctuating_2024}, 
%(i) we take an order $N =2$ truncation and then 
we take (i) $ \partial_t a_{\pm 1} =0$ as we suppose coefficients of order $n>1$ relax ``quickly'', and (ii) over large length scales terms of order $\mathcal{O}(\nabla^2_r)$ can be neglected, as they sense small length scale variations -- the so called ``diffusion-drift approximation''. Looking back on Eq.~\eqref{eq:coefficients} with these approximations we find
\begin{align*}
    \partial_ta_0(\boldsymbol{r},t)&=-\frac{v}{2}\nabla_r\cdot \left[ \hat{i}_+a_{1}+\hat{i}_- a_{-1}\right] + D_t\Delta a_0 \\
    0&=-\frac{v}{2}\nabla_r\cdot \left[ \hat{i}_\mp a_{0}\right] - (D_r+\alpha) a_{\pm1} + \mathcal{O}(\nabla^2_r)
\end{align*}
which we can easily solve for to find a partial differential equation for $a_0$ only, 
\begin{equation*}
    \partial_ta_0(\boldsymbol{r},t) = D_{\rm eff}\nabla^2a_0
\end{equation*}
where we recall that $D_{\rm eff} = \left( \dfrac{v^2}{2(D_r + \alpha) } +D_t \right)$.
This is the evolution equation of a diffusive process with effective diffusion coefficient $D_{\rm eff}$. In this limit, the process is once again Gaussian and we recover Eq.~\eqref{eq: diffusive regime} with $D=D_{\rm eff}$. Diffusive regimes behave similarly for our Gaussian and non Gaussian models. This is likely because the non-gaussianity in RTP and ABP comes from the self-propulsion part of the dynamics which dominate only at intermediate time scales.  %\tristan{tu penses que c'est vrai cette phrase?} \tristan{Je dirais juste garder l'idée de self-propulsion parce que ça vient du fait que sa norme soit constante, et la reorientation est importante at long times aussi}

To check the relevance of these limiting diffusive laws, we present in Fig.~\ref{fig:All NMSD}-b, where time is rescaled by a diffusive time scale, the limit regime of Eq.~\eqref{eq: diffusive regime} with $D = D_t$ and $D = D_{\rm eff}$. The curves indeed all collapse at short time scales onto the limiting law with $D_t$ and at long timescales onto that with $D_{\rm eff}$ confirming our analysis. 

\begin{figure}
    \centering
    \includegraphics[width=0.99\linewidth]{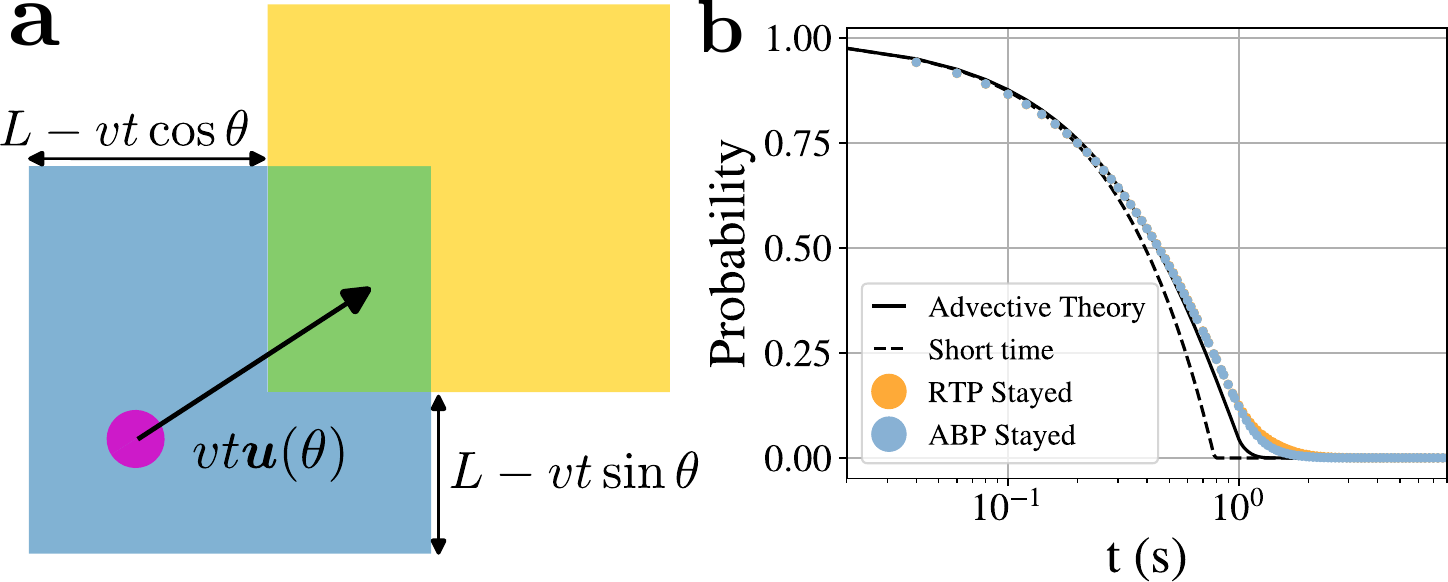}
    \caption{\textbf{Probability distribution from geometric arguments.} Schematics (a) illustrating the probability distribution clouds to find an advective particle in a box after some time $t$ (yellow) given it started in the box initially (blue). Probability distribution of ABPs or RTPs to remain in the box they start in (b) obtained from simulation against ones obtained from geometric arguments in the advective case. Simulation parameters are the same as in Fig.\ref{fig:ISF_ABP_RT} and we consider boxes of size $L = 5~\unit{\mu m}$.}
    \label{fig:proba}
\end{figure}

\paragraph{Intermediate time scales}
Finding an expression for the advective regime of ABPs and RTPs requires more work. To do so, one option could be to neglect translational diffusion, taking $D_t =0$ in Eq.~\eqref{eq:a0 order1}, and inquire then about short time behaviour. However, this strategy is not sufficient as still the integrals in Eq.~\eqref{eq:Cnt} cannot be solved. To circumvent this issue, we resort to the other expression of $C_N(t)  = \langle N \rangle P_{\rm in}(t)$ given in Eq.~\eqref{eq:probability}. Clearly, to obtain approximate predictions for $C_N(t)$, it is sufficient to search for approximate expressions for $P_{\rm in}(t)$, the probability for a particle to still be in the box after some time $t$ knowing it was in the box initially.

To adress the intermediate time scales regime, we can consider that at these timescales, particles move with speed $v\boldsymbol{u}(\theta)$, and do not yet reorient ($D_r = \alpha =0$). We also consider they do not diffuse ($D_t =0$). Consider such a particle starting anywhere in a box at $t=0$ -- then, its probability distribution cloud can be represented as a blue square -- as sketched in Fig.~\ref{fig:proba}-a. At some later time $t$, this probability distribution is displaced by a vector $v t \boldsymbol{u}(\theta)$, as represented by the yellow square. The probability of the particle of still being in the box at time $t$ corresponds to the overlap between the blue box and the yellow box. We can then write the probability $p_{\rm in} (\theta,t)$ of the particle with orientation $\theta$ of still being in the box as
\begin{equation}
\label{eq:full proba advective}
    p_{\rm in} (\theta,t) =
    \begin{cases}
        \frac{(L-v|\cos{\theta}|t)(L-v|\sin{\theta}|t)}{L^2} \\
        0 \, \, \text{if } v\cos{\theta}t \geq L \, \, \text{or} \, \, v\sin{\theta}t \geq L.
    \end{cases}
\end{equation}
To unveil the limiting intermediate regime we restrict ourselves to short enough times in the expression above such that $p_{\rm in}(\theta,t)$ is non zero, and neglect terms of order $\mathcal{O}(t^2) $. %\tristan{explain that we do this to look for limiting regime of intermediate behaviour}

We can then average $p_{\rm in}(\theta,t)$ over all the possible orientations $\theta$ to obtain at short times
\begin{equation}
\label{eq:Angular average Proba advective}
    P_{\rm in}(t) = \frac{1}{2\pi} \int_0^{2\pi} \, \dd\theta \,p_{\rm in} (\theta,t) \simeq 1 - \frac{4}{\pi}\frac{v}{L}t.
\end{equation}
As an intermediate validation, we check that this equation accurately reproduces the probability that either an ABPs or an RTPs is still in a box at short enough times in the intermediate regime, as obtained from simulations Fig.~\eqref{fig:proba}-b. The short time approximation in Eq.~\eqref{eq:Angular average Proba advective} captures behavior accurately at short enough times (black dashed line). Averaging over $\theta$ the ``all time'' expression of Eq.~\eqref{eq:full proba advective} numerically (black full line) pushes the agreement to longer timescales. At long enough times, $P_{\rm in}(t)$ is sensitive to reorientation dynamics and the probability that ABPs and RTPs are still in a box is, accordingly, no longer well captured by our approximations.

The probability $P_{\rm in}(t)$ then gives $C_N(t)$ and the NMSD
%&= 2(\langle N^2 \rangle - \langle N \rangle^2) - 2C_N(t) \notag \\
    %&=2\left(\langle N \rangle - \langle N \rangle \left(1-\frac{4}{\pi}\frac{v}{L}t\right) \right) 
\begin{align}
    \langle \Delta N(t)^2\rangle \simeq \frac{8}{\pi}\langle N\rangle\frac{v}{L}t \label{eq:advective regime ABP/RT}
\end{align}
which is valid at short time scales. Eq.~\eqref{eq:advective regime ABP/RT} is presented in Fig.~\ref{fig:All NMSD}-c, and agrees perfectly with the full theory and simulation data at intermediate regimes. We thus obtain a very similar law as for Eq.~\eqref{eq:advective regime AOUP}, but with a prefactor that is higher compared to the AOUP case.  

Why does this difference arise? Our probabilistic method for $P_{\rm in}(t)$ can also be used to investigate AOUPs, but in that case, one has to realize that the probability distribution $p_{\rm in}(\theta, |\vaoup|, t)$  of a particle with orientation $\theta$ to still be in the box at time $t$ also depends on the norm of the velocity $|\vaoup|$, which is itself a distribution. Clearly the same reasoning as in Eq.~\eqref{eq:full proba advective} still applies, but then one needs to average over all possible values of $\theta$ and $ |\vaoup|$.
Each component of the speed vector $\vaoup$ of an AOUP is Gaussian distributed, thus the magnitude $|\vaoup|$ follows a Rayleigh distribution $P(|\vaoup|) = \frac{|\vaoup|\tau_r}{D_v}e^{-\frac{|\vaoup|^2 \tau_r}{4D_v}}$. One then gets
%\langle \Delta N(t)^2\rangle &= \frac{8}{\pi}\langle N\rangle\frac{t}{L} \int_0^{\infty}  |\boldsymbol{v}| P(|\boldsymbol{v}|) \, \dd |\boldsymbol{v}| \\&= \frac{4}{\sqrt{\pi}} \langle N \rangle \sqrt{\frac{2D_v}{\tau}}\frac{t}{L}
\begin{align*}
    P_{\rm in}(t) &=  \frac{1}{2\pi} \int \dd |\vaoup| P(|\vaoup|) \int_0^{2\pi} \, \dd\theta  \,p_{\rm in} (\theta,|\vaoup|,t) \\
    &\simeq 1 - \frac{2}{\sqrt{\pi}}\frac{\sqrt{2D_v/\tau_r}}{L}t.
\end{align*}
Identifying $\sqrt{2D_v/\tau_r} = v$ as the mean velocity of an AOUP, and inserting this expression for $P_{\rm in}(t)$ in the NMSD, one recovers Eq.~\eqref{eq:advective regime AOUP}.
The differences in the distribution of self propulsion speeds thus translates to different prefactors in the advective, intermediate time regime of the Countoscope. The difference in prefactors is small, of about $10\%$, since $4/\sqrt{\pi} \simeq 2.25$ and $8/\pi\simeq 2.55$, however it is present. The non gaussianity of the ABP and RTP models is thus seen in the advective regime, and compared to a gaussian model, differences in the NMSD originate from a peaked velocity distribution rather than from reorientation dynamics. In our companion paper, we find that differences in reorientation dynamics are more easily seen in the correlation function $C_N(t)$~\cite{cerdin2026number}.

% \tristan{We already talked about this in the PRL, can we actually just remove all of that discussion of timescales?}
% One direct consequence of this difference between models is the transition time between each regimes. While for AOUPs using the expressions we derived we have for $t_{adv}$ the time it takes to reach the advective regime :
% \begin{align*}
%     \frac{8}{\sqrt{\pi}}\langle N \rangle \sqrt{\frac{D_t t_{adv}}{L^2}} &=  \frac{4}{\sqrt{\pi}}\langle N \rangle \sqrt{\frac{2D_v}{\tau}}\frac{t_{adv}}{L} \\
%     t_{adv}&=4 \frac{Dt \tau}{2D_v}=4\frac{D_t}{v^2}
% \end{align*}
% And similarly we obtain for the time to reach long time diffusion $t_{diff} = 4\frac{D_{eff}}{v^2}$. These are equal to the times to transition between regimes in the MSD. 
% However the result differs for ABPs and RTPs. We have :
% \begin{align*}
%     \frac{8}{\sqrt{\pi}}\langle N \rangle \sqrt{\frac{D_t t_{adv}}{L^2}} &= \frac{8}{\pi}\langle N\rangle\frac{v}{L}t_{adv} \\
%     t_{adv}&=\pi \frac{D_t}{v^2}
% \end{align*}
% And similarly $t_{diff} = \pi \frac{D_{eff}}{v^2}$. Thus for the number fluctuations of ABPs and RTPs the transition between regimes happens faster than in the MSD. This is of importance since a common challenge in the acquisition of experimental data of active particles is to observe them over time scales long enough to be able to characterize properly the diffusive regime. Being able to observe it earlier in the NMSD is thus advantageous. 

All the limiting regimes that we have obtained %can be seen in the panels b anc c of Fig.~\ref{fig:All NMSD}. They perfectly describe the sections where all the curves collapse onto one under the effect of our rescaling. We have thus obtained a set of 
provide simple analytical descriptions linking the behavior of each regime with the dynamic parameters that are relevant to quantify it ($v$, $D_t$, $D_{\rm eff}$). These laws provide one way to quantify the motion of particles, using simple fits on number fluctuations calculated from experimental data. 

\begin{figure*}
    \centering
\includegraphics[width=0.99\linewidth]{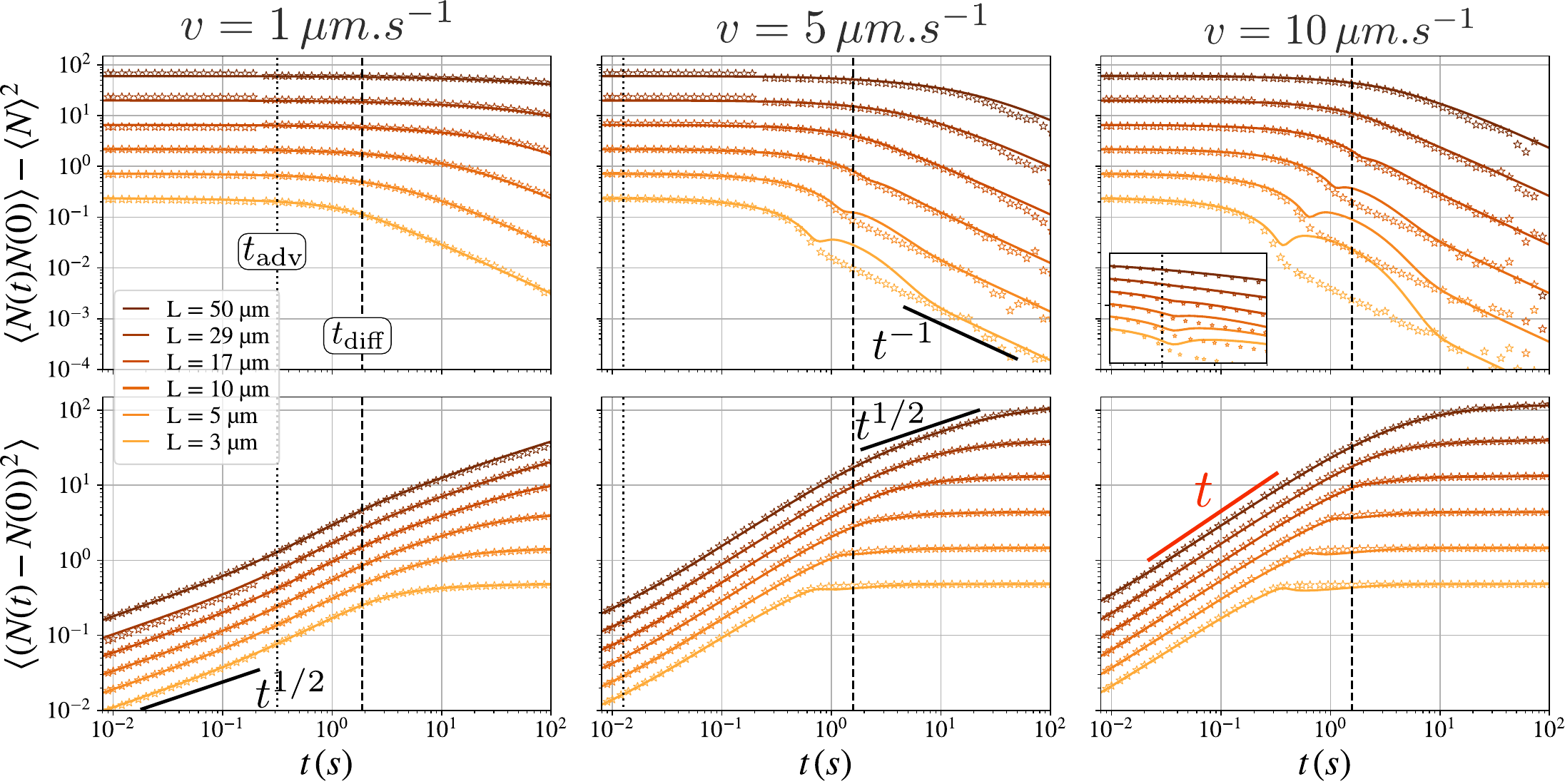}
    \caption{\textbf{Number fluctuations for Run-and-tumble particles at different Péclet numbers.} (Top row) Number correlations $C_N(t)$ and (bottom row) NMSD, with $v$ increasing from left to right, $v = \{1,5,10\}~\mathrm{\mu m~s^{-1}}$ corresponding to $\text{Pe} \simeq\{ 1.6,8,16\}$. All other dynamic parameters are kept constant at the same values as in Fig.~\ref{fig:All NMSD}. Stars correspond to simulated data and colored lines to the theory truncated at $N = 2$, using Eq.~\eqref{eq:Cnt} and Eq.~\eqref{eq:f2kt}. The inset in the rightmost column shows $C_N(t)$ with time rescaled by an advective timescale $v/L$ and a vertical black dotted line at $t = v/L$. }
    \label{fig:Péclet}
\end{figure*}

\subsection{Convergence at different orders for Countoscope metrics}

It is known from study on the MSD of active particles that fitting the full expression should be preferred to fitting limiting regimes to obtain more accurate values of the dynamic parameters~\cite{bailey2022fitting}. In this context, it is important to know at which order in the truncation one should stop to obtain accurate enough results for the NMSD. 

\paragraph{Convergence with respect to the $\text{Pe}$ number.} An important parameter affecting the accuracy of the truncation is the value of the Péclet number. This is expected from the construction of our $\widetilde{a}_0(\boldsymbol{k},\omega)$ as highlighted in Sec.~\ref{sec:hydro demonstration}, given that the value of the continued fraction terms $\frac{c_n}{b_n}$ is larger at higher $\text{Pe}$ numbers. Several authors have previously highlighted that typical truncation schemes lose precision at higher $\text{Pe}$~\cite{broker_pair-distribution_2023,gautry_closures_2025}.
Focusing on order $N = 2$ truncation, the highest order for which we obtained analytical results for $F(\kk,t)$, we plot NMSD and $C_N(t)$ curves for RTPs at increasing $\text{Pe}$ in Fig.~\ref{fig:Péclet}. For this investigation we focus on RTPs, since we know RTP approximations converge more slowly than for ABP. By keeping all dynamic parameters the same, except for the self propulsion speed that takes values $v = \{1,5,10\}$, we probe $\text{Pe} \simeq\{ 1.6,8,16\}$, covering the range of most experimental systems. 

For the lowest values of $\text{Pe}$, the particles' behaviour is close to that of passive particles with the advective regime barely appearing in the NMSD. Our analytical expressions are accurate in this situation for both $C_N(t)$ and the NMSD. At intermediate $\text{Pe}$ value, deviations between the theory and simulation data appear, especially in the number correlations $C_N(t)$ for the smallest boxes. This phenomenon is more pronounced at higher $\text{Pe}$ value, and visible in bigger boxes. Generally, the disagreement starts at timescales of order $t=L/v$, as seen in the inset of Fig~\ref{fig:Péclet}. Beyond this timescale, particles have typically traversed the box via advection, and contributions to $P_{\rm in}(t)$ are dominated by reorientation dynamics which determine whether or not a particle returns to a box it has left. Naturally, this effect is more pronounced in small boxes that particles are more likely to cross via advection without reorienting while they cross through it, and, therefore, the truncated theory fails to correctly predict behavior on smaller boxes. This behavior can be described through a critical lengthscale, $L_c \sim t_{\rm diff}v = \dfrac{\pi D_{\rm eff}}{v}$, which characterizes the typical maximum value of box sizes for which these effects should be seen. We can calculate that $L_c = \{ 1.9, 7.9, 15.7\}~\mathrm{\mu m}$ for our values of $v$. We find indeed that so long as $L \geq L_c$, the theory is in excellent agreement with the simulations. 

Interestingly the departure from the theory in the correlation functions $C_N(t)$ are not as apparent in the NMSD (bottom row of Fig.~\ref{fig:Péclet}). In fact, deviations occur primarily at long times in the correlation function, and their amplitude is small compared to the value of $C_N(t=0) = \langle N^2 \rangle - \langle N \rangle^2$. In the NMSD space, this corresponds to small differences between the NMSD at late times and its plateau, and deviations between the theory and simulations are effectively hidden. The most noticeable differences occur only for the smallest box sizes (yellow) around the time where the NMSD reaches its plateau, at the highest \text{Pe} values. There the theory exhibits an unphysical albeit very little hump. Therefore, the visible inaccuracy in $C_N(t)$ has limited impact on the NMSD. The truncation at order $N = 2$ presents only noticeable disagreements in the NMSD for the highest $\text{Pe}$ values and smallest box sizes, which are close to that which can be observed in the motion of low density bacteria.  

\begin{figure*}
    \centering
    \includegraphics[width=0.99\linewidth]{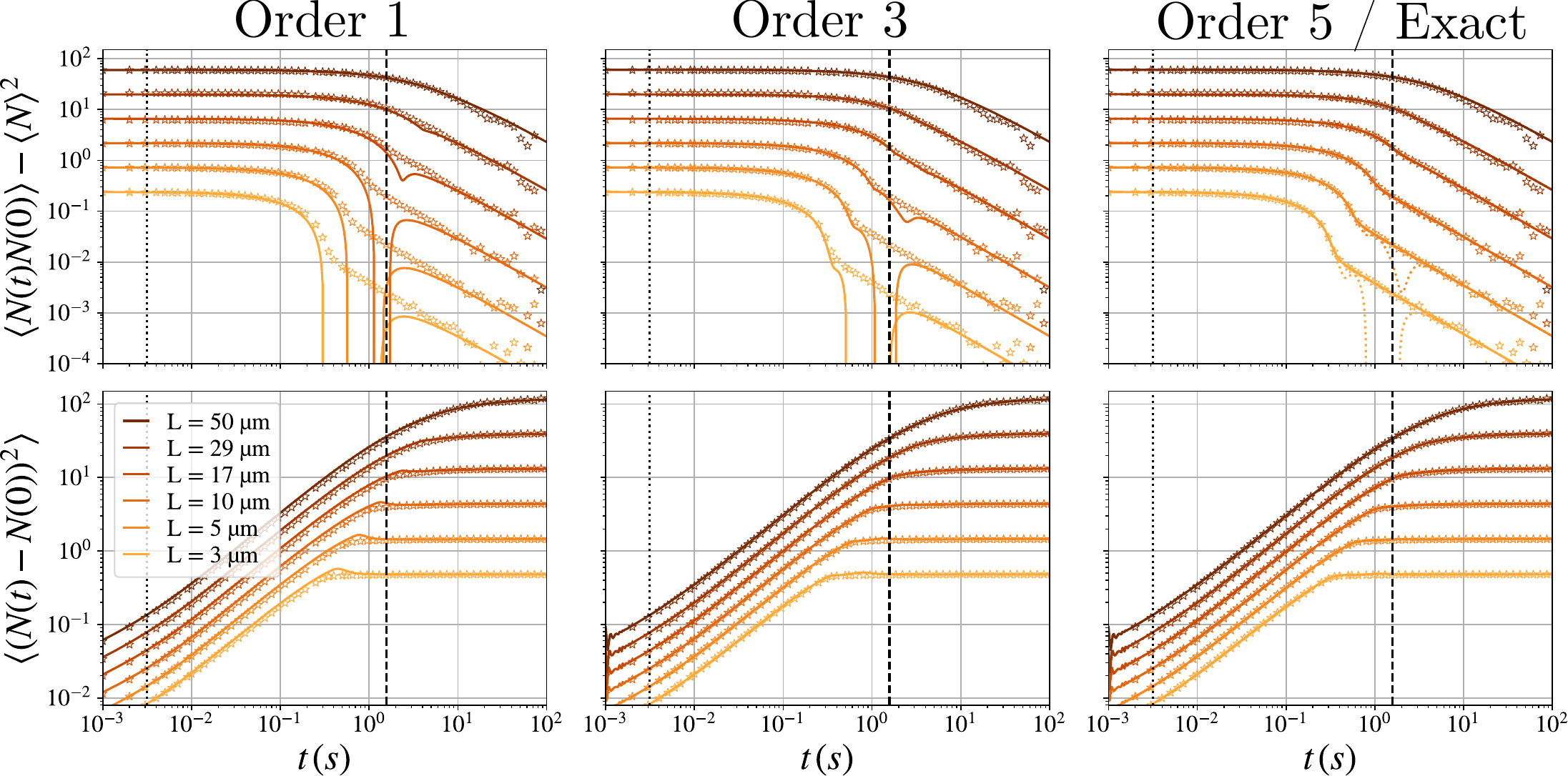}
    \caption{\textbf{Number correlations and NMSD for Run-and-tumble particles at different Orders} in the truncation. Orders are increasing from left to right, with the rightmost column showing the exact expression in full lines and the Order 5 result in dotted lines.}
    \label{fig:RTP Order Countoscope}
\end{figure*}

\paragraph{Convergence with respect to the order $N$ of the truncation scheme.} We now restrict ourselves to investigations at high $\text{Pe}$ number where disagreement between approximate theory and simulations is the most apparent, \textit{e.g.} at a high self-propulsion speed $v=10~ \unit{\mu m~s^{-1}}$. We then look at how this disagreement can be reduced at increasingly higher truncation orders $N$ in Fig~\ref{fig:RTP Order Countoscope} (see also Fig.~\ref{fig:Cnt pair orders}). 

At order  $N=1$ (left column of Fig~\ref{fig:RTP Order Countoscope}) the theoretical approximation is, as expected, less accurate than for $N = 2$ that was investigated in Fig.~\ref{fig:Péclet} (right column). Differences between the theory and simulations are more striking and can even be seen clearly in the NMSD, at the time where the NMSD reaches its plateau, for several box sizes. In addition, at order 1 the prediction for the correlation function is that it becomes negative, which translates into a huge drops in log-log scale on smaller boxes. Because of an odd-even periodicity, these drops are typically only observed for odd truncation orders $N$, and not for even $N$. 

At order $N =3$ the accuracy is already noticeably better. Values of the correlation function still appear to become negative for some boxes, but over a reduced range time range, and box size range. %Therefore some values of $C_N(t)$ that disagree with simulated ones can become negative, leading to what seems like a divergence in log-log scale. However one can see on boxes big enough were this effect doesn't occur and small enough for the theory to still present some errors that at order 3 the theory is more accurate. 
Accordingly, the NMSD presents a seemingly perfect agreement between theory and simulation for all observed box sizes. 
Order $N=5$ is even more accurate, with some minor differences remaining in $C_N(t)$, and none for the NMSD. The exact formula in Eq.~\eqref{eq:a_0 RTPs} shows excellent agreement at all time and length scales. 
Similar conclusions at all orders are also true for ABPs (see Fig.~\ref{fig:ABP Order Countoscope}).

This investigation allows us to draw some intermediate conclusions on model choosing.
At the NMSD level, there are barely any noticeable differences between orders $N = 3, 5$, and the exact expansion. It is thus natural to use $N = 2$ or $N=3$ to capture the behavior of the NMSD in a practical application. 
For the correlation function however, higher order approximations are necessary. While an order 3-5 is sufficient for ABPs (see Fig.~\ref{fig:ABP Order Countoscope}), more orders are required to accurately reproduce RTPs, and hence one should use the exact formulation -- which comes at no (if not at reduced) additional computational cost. 
%There are however no reason to go to order 5 in the truncation instead of using the exact value, as the only demerit of the exact expression is the need to numerically integrate it in order to obtain the ISF. 
%\tristan{If they are easy to do, I would add orders $2/4/6$ for just $C_N(t)$ in the SI and reference them somewhere around here.} .

\begin{figure}
    \centering
    \includegraphics[width=0.99\linewidth]{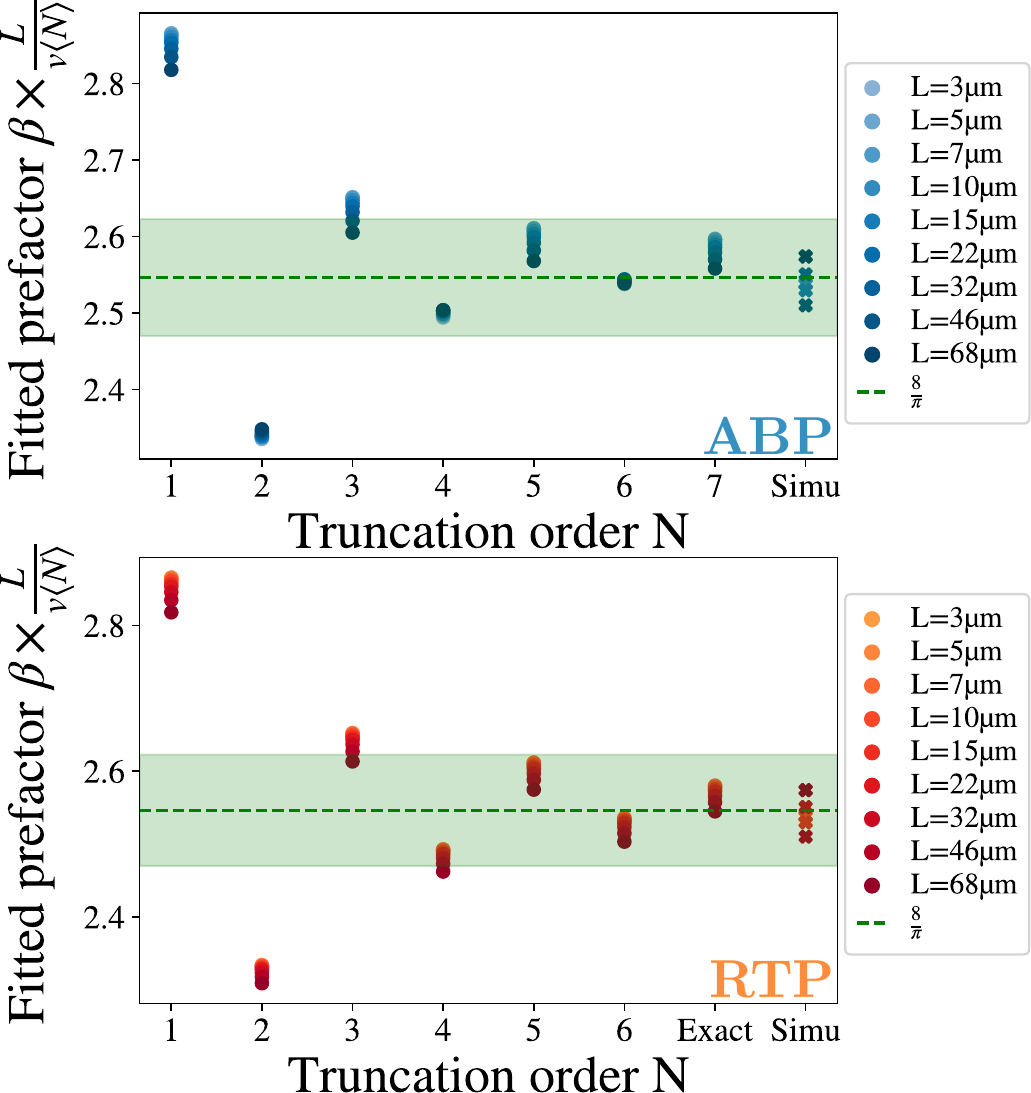}
    \caption{\textbf{Value of the prefactor to the advective regime} for ABPs (left) and RTPs (right) obtained from fitting the theory at different orders compared to the theoretical value of $\dfrac{8}{\pi}$.}
    \label{fig:Adv_coeff}
\end{figure}

\paragraph{Convergence of the limiting regimes}
We can quantify this increasing accuracy by investigating the limiting regime at intermediate timescales. The exact behavior at intermediate timescales is given by Eq.~\eqref{eq:advective regime ABP/RT}, and our aim is to quantify how fast with increasing $N$ we can capture this exact prefactor. To do so, we investigate a system at high $\text{Pe}$ number. We then fit a linear law at short times, $\langle \Delta N^2(t)\rangle = \beta t$, where $\beta$ is a fitting parameter. We then compare $\beta$ with our theoretical expression in Eq.~\eqref{eq:advective regime ABP/RT}, and plot the results with increasing truncation order $N$ in Fig.~\ref{fig:Adv_coeff}. Given we plot $\beta L/v \langle N \rangle$, we expect this prefactor converges to $8/\pi$.

The results obtained clearly approach the $8/\pi$ theoretical prediction with increasing truncation order $N$.  Using the exact expression for the ISF of RTPs Eq.~\eqref{eq:a_0 RTPs}, we get as close to the correct value as our fit precision allows. Overall, the prefactor $\beta$ oscillates with increasing order $N$, as odd orders overestimate the coefficient $\beta$ while even orders underestimate it (see Fig.~\ref{fig:Cnt pair orders} for even values of $N$). This is intimately linked to the behaviour observed in the curves of $C_N(t)$ where an upward or downward bump appeared compared to simulations for even and odd orders respectively. 

%\sophie{I think we can remove that paragraph. I am ready to believe this exists and don't really care about it at this stage but we can leave it in SI.} 
This can be explained from the structure of our continued fractions. Indeed we can rewrite through an equivalence transformation
\begin{equation*}
    \widetilde{a}_0^{-1} = b_0 + \K_{n=1}^N \frac{c}{b_n} = b_0 + \K_{n=1}^N \frac{1}{d_nb_n}
\end{equation*}
with\begin{align*}
    d_{2n} = 1 , \,\,\text{and} \,\,  d_{2n +1} = \frac{1}{c},\,\, \forall n \in \mathbb{N}.
\end{align*} 
This explains our oscillating behaviour, as stopping at an even order will give us a smaller continued fraction, thus a bigger $\widetilde{a}_0$ and we will overestimate $C_N(t)$, and the opposite happens stopping at an odd order. 

At any order we notice a degree of uncertainty around a mean value depending on the value of the box size used. Investigating fits of the NMSD obtained from simulations, knowing the self-propulsion velocity $v$, shows that the value of the prefactor is roughly $\pm 2\%$ away from the expected prefactor depending on the box size (see the last point on the x axis ``Simu'' in Fig.~\ref{fig:Adv_coeff}). We can conclude that a roughly $\pm 2\%$ error is only a fitting error and that orders from $N\geq 3$ should provide satisfactory estimates of the self-propulsion speed. 

% \sophie{actually -- if you were to fit experimental data that would be a different and reverse thing. What happens if you fit a line to your numerical data? (using the known $8/\pi$ prefactor)? Are you that far away from the ``exact result''. I think you should do that test and then we decide what to say. For instance you do it for a few box sizes and average the results}
We conclude by tempering the results of the fit in Fig~\ref{fig:Adv_coeff} by insisting that in order to correctly estimate dynamic coefficients one should fit the full expression instead of limiting regimes. We anticipate that the accuracy of order 3 for the NMSD should be enough in most situations to justify using it, as it admits an analytic result for the ISF and is thus easier to obtain for a wide range of values. Precise discussion on the use of our expressions at different orders to fit experimental data is however outside the scope of this paper and is left for future work.

\section{Conclusion}

We have here established theoretical expressions to quantify number fluctuations of three canonical models of self-propelled particles: Active Brownian (ABPs), Run and Tumble (RTPs) and Active Ornstein Uhlenbeck (AOUPs). For all models, we relate number fluctuations to an integral of the ISFs through Eq.~\eqref{eq:Cnt}. AOUPs have gaussian statistics and hence can be treated generically under a gaussian framework; giving an ISF which can easily be integrated to give an explicit analytical expression Eq.~\eqref{eq:Cnt Gaussian} for the number correlations. For ABPs and RTPs, we have provided a novel derivation of the ISFs, allowing us to get both exact or approximate expressions -- exact expressions are usually more precise but longer to evaluate numerically. 
Interestingly our derivation bridges together the methods consisting in solving the system through propagators and successive reorientation steps~\cite{zhao_quantitative_2024,martens_probability_2012} or doing an expansion of the FPEs on the angular order of the perturbations to the density. The approach allows us to gain physical insights into the features of the ISFs and the number correlations: namely that oscillations originate from reorientation dynamics. All our theoretical expressions allow us to uncover 3 limiting regimes of the mean squared number displacement in time: at short times diffusive, at intermediate times ballistic, and at long times diffusive with enhanced diffusion. 

How might we use these expressions in practice to fit experimental data? Since this question deserves a self-contained investigation, we reserve this for a later work and rather discuss general guidelines. Since we have a few analytical laws that are  straightforward to calculate -- such as the limiting regimes Eq.~\eqref{eq:advective regime ABP/RT} and Eq.~\eqref{eq: diffusive regime} -- these can be used to obtain a range of relevant dynamical parameters of the particles. The integrated ISFs in Eq.~\eqref{eq:Cnt} using the approximate ISFs of Eq.~\eqref{eq:a0 order1} or Eq.~\eqref{eq:a0 order2} can serve to refine fitting, provided the P\'eclet number of the particles is small enough, and if not, then the exact expressions, either Eq.~\eqref{eq:a_0 RTPs} for RTPs or Eq.~\eqref{eq:a_0 continued fraction full} for ABPs should be used.

%\tristan{Un autre commentaire c'est est-ce qu'on peut utiliser le formalisme pour d'autres reorientation dynamics?? par exemple run-reverse?}

Here, we conducted derivations for 2D systems and so our results are not straightforwardly applicable to \textit{e.g.} bacteria suspensions that are often found swimming in 3D~\cite{kurzthaler2024characterization}. However, similar formalisms may be used to obtain expressions for the ISFs in 3D~\cite{martens_probability_2012,gautry_closures_2025} and so one can hope to use them to predict behavior of number fluctuations. While quantitatively these results might differ slightly, qualitatively we expect all features uncovered here to be consistently observed in 3D. Indeed, gaussian models like AOUPs are similarly expressed in 3D as in 2D, with a slight modification of Eq.~\eqref{eq:Cnt Gaussian} into
\begin{equation}
    C_N(t) = \langle N \rangle \left[ f\left( \frac{\langle \Delta r^2(t) \rangle}{L^2} \right) \right]^d
\end{equation}
where $d$ is the system's dimension. Therefore, we can expect all limiting regimes of the mean squared displacement to arise in the NMSD. Similarly, the derivations which provide the limiting laws Eq.~\eqref{eq:advective regime ABP/RT} and Eq.~\eqref{eq: diffusive regime} can easily be extended to 3D. %In particular for ABPs and RTPs the intermediate regime would be given by (details not shown here)
%\begin{equation}
%    \langle \Delta N^2(t) \rangle \simeq 3 \langle N \rangle \frac{vt}{L}
%\end{equation}
%similar to Eq.~\eqref{eq:advective regime ABP/RT} but with a slightly different prefactor. 
We can thus anticipate the use of the technique also in systems with 3D dynamics. 

While we have focused here on non-interacting particle systems, it is natural to ask what kinds of behaviors would be observed at higher densities. Since real space number correlations provide easily physically interpretable signals, then we might expect they would also help to shed some light on collective behavior of active suspensions. Including a density dependent self propulsion for instance could be done following \cite{dinelli_fluctuating_2024}.
Remarkably, the investigation of \textit{static} number fluctuations $N$ in observation volumes has proven useful in quantifying \textit{static} properties of dense non-equilibrium systems for over 30 years. For instance
``giant'' number fluctuations, where $\alpha > 0$ in the scaling $\langle N ^2 \rangle - \langle N \rangle^2 \sim N^{1 + \alpha}$, indicate long-range organization in bacterial or synthetic
active matter suspensions~\cite{zhang2010collective,peruani2012collective,liu2021density,fily2012athermal,dey2012spatial,alarcon2017morphology,chate2008collective,chepizhko2021revisiting,fadda2023interplay,narayan2007long,toner2005hydrodynamics,navarro2015clustering}.  It is thus clear that investigating \textit{dynamic} number fluctuations in virtual observation boxes like $N(t)$, is a promising route to quantify collective dynamic features of self-propelling suspensions.

\section*{Acknowledgements}

The authors acknowledge many fruitful discussions for the early elaboration of this work especially with Federico Paratore but also with Laura Alvarez, Lucio Isa, Ueli T\"{o}pfer and Carolijn van Baalen. Naoufal Elaisati was especially helpful as he established some preliminary work on this topic. Further discussions with Maxime D\'efor\^et, Timoth\'ee Gautry, Pierre Illien, Rapha\"el Jeanneret, Christina Kurzthaler, Arnold Mathijssen were also helpful. We are also grateful to Adam Carter for sharing his code for making intermediate scattering functions. 

\section*{Author contribution statement}

Conceptualization: T. Cerdin, S. Marbach; Methodology: T. Cerdin, S. Marbach, C. Douarche; Formal Analysis: T. Cerdin; Overall investigation: T. Cerdin, S. Marbach; Investigation of Limiting Regimes: T. Cerdin, T. Calazans; Visualization: T. Cerdin; Supervision: C. Douarche, S. Marbach, Writing – Original Draft: T. Cerdin, S. Marbach, C. Douarche;  Writing – Review \& Editing: T. Cerdin, S. Marbach, C. Douarche;

\appendix

\section{Computational Methods}
\label{appendix:computational methods}

For all three active particle models we simulate $1500$ particles within a square simulation box of size $L_{\rm box} = 250 ~\unit{\mu m}$ and periodic boundary conditions. As highlighted in~\cite{carter2025measuring}, the periodic boundary conditions restrict the maximum size of boxes to be used for counting to about a third of the window size -- otherwise periodic effects start to be seen. The dynamics are resolved using an Euler-Maruyama scheme to numerically integrate the stochastic equations of motion. In the case of RTPs we perform the random reorientation with a rate $\alpha$ using a Gillespie algorithm. To resolve the dynamics with sufficient accuracy over 5 orders of magnitude in time, we perform both short simulations with a very small time step $\mathrm{d}t = 0.001s$ which is equal to the time step and longer simulations with time step $\mathrm{d}t = 0.005$ while saving every $0.05 s$. We then merge the results together on plots for $C_N(t)$ and $\langle \Delta N^2(t) \rangle$. 

The ISF of simulations is computed on the long simulated datasets using the python library~\cite{carter_2025_17362168} based on the direct method of calculation. 

Numerical integration of the ISFs of Order 3 and above are done using an inverse FFT with a custom routine in python. They are then saved for every set of dynamic parameters to be used to compute $C_N(t)$ at any box size.

Numerical integration of the $\boldsymbol{k}$ integrals in Eq. \ref{eq:Cnt} for ABPs and RTPs are done following the scheme outlined in~\cite{mackay2024countoscope}. We carry out the $\boldsymbol{r}$ integrals :
\begin{align}
\label{eq:c_n k int}
     C_N(t)=  \langle N \rangle\int  \frac{d\boldsymbol{k}}{(2\pi)^2} L^2 \prod_{i = x,y}\left( \frac{\sin(\frac{k_iL}{2})}{k_i L/2} \right)^2  F(\boldsymbol{k},t).
\end{align}
To facilitate the numerical integration, we change coordinates from cartesian $\{k_x L/2,ky L/2\}$ to rescaled spherical coordinates $\{ K, \phi\}$. $C_N(t)$ can then be re-expressed as 
\begin{equation}
\label{eq:cnt spherical}
    C_N(t)=  \langle N \rangle\int  \frac{KdK}{\pi^2} f_v(K) F(2K/L,t)
\end{equation}
with :
\begin{equation}
    f_v(K) =\int_0^{2\pi} d\phi\left( \frac{\sin(K\cos{\phi})}{K \cos{\phi}} \right)^2 \left( \frac{\sin(K\sin{\phi})}{K\sin{\phi}} \right)^2.
\end{equation}
This separates the volumetric term in front of the integrals, which is quickly oscillating and complicates the numerical integration. As it is constant it can be computed once precisely and then reused subsequently in all Countoscope computations. Using it alongside the obtained values of $F(\boldsymbol{k},t)$ is how we speed up the final numerical integration. 

The probability of particles to stay or leave boxes is found by computing for each lag time $\tau$ the probabilities $P_{\rm stayed}(\tau)$ (particle remains continuously inside) and $P_{\rm returned}(\tau)$ (particle exits then re-enters and is present at time $\tau$), averaged over all boxes and initial times.

All counting on simulated data to obtain $C_N(t)$ and the NMSD was done using the Countoscope code published in \cite{sophie_marbach_2025_15000583}.
Upon publication, code for this specific investigation will be added on the same repository. %\tristan{That's the plan right ? How do I cite it ?}

\section{Equivalence between harmonic coefficients and hydrodynamic fields}
\label{appendix:equivalence}

To solve Eq.~\eqref{eq:FPE}, an alternative method~\cite{solon2015active, gautry_closures_2025} is to use another decomposition of the probability distribution function (PDF) $P(\boldsymbol{r},\theta,t)$. Instead of using harmonic coefficients $a_n$, one can  decompose the PDF in terms of hydrodynamic fields as 
\begin{equation}
    P(\boldsymbol{r},\theta,t) = \rho(\boldsymbol{r},t) + \boldsymbol{p}(\boldsymbol{r},t)\cdot \boldsymbol{u}(\theta) + \underline{\underline{q}}(\boldsymbol{r},t): \left( \boldsymbol{u}  \boldsymbol{u}-\frac{\mathbb{I}}{2} \right).
\end{equation}
Writing out explicitly this expansion against the one we used (Eq.~\ref{eq:decomposition harmonics}) we have 
\begin{align}
    P(\boldsymbol{r},\theta,t) &= \rho +  p_x \cos \theta + p_y  \sin \theta \\ & + \frac{1}{2} [(q_{xx}+q_{yy})\cos{2\theta} + (q_{xy} + q_{yx})\sin{2\theta}] \notag \\ 
    &=a_0 + e^{i\theta}a_1+ e^{-i\theta}a_{-1} + e^{2i\theta}a_2 + e^{-2i\theta}a_{-2} \\
    &= a_0 +  (a_1+a_{-1}) \cos \theta + i(a_1-a_{-1})  \sin \theta 
    \\&+ [(a_2 + a_{-2})\cos{2\theta} + i(a_{2}-a_{-2})\sin{2\theta}]. \notag
\end{align}
From this we immediately deduce that we can express 
\begin{align}
    \rho(\boldsymbol{r},t) &= a_0(\boldsymbol{r},t)\\
    \boldsymbol{p}(\boldsymbol{r},t) & = \begin{pmatrix}
        a_1+a_{-1} \\ i(a_1 - a_{-1}) 
    \end{pmatrix}.
\end{align}
Since $\underline{\underline{q}}$ is chosen to be a symmetric traceless matrix we can find that $q_{xy} = q_{yx} = i(a_2 + a_{-2})$. For the diagonal components of $\underline{\underline{q}}$ we can compare Eq.~\eqref{eq:coefficients 2} with the evolution equation obtained in~\cite{solon2015active} 
\begin{equation*}
    \partial_t{q}_{ij} = -2 B_{ijkl}\partial_l(vp_l) - (4D_r+\alpha)Q_{ij} +Dt\Delta Q_{ij} - \nabla_k \chi_{ijk}
\end{equation*}
where $B_{ijkl} = \frac{1}{4}(\delta_{ik}\delta_{jl} + \delta_{il}\delta_{jk} - \delta_{ij}\delta_{kl})$ and summation over repeated indices are implied. 
Writing it out for the diagonal components and ignoring harmonic contributions of order 3 and above contained in $\chi_{ijk}$ we have 
\begin{align}
    \partial_t q_{xx}&= -\frac{v}{4} \nabla \cdot \left[ \begin{pmatrix} 1 \\ -1\end{pmatrix} \boldsymbol{p} \right]+ D_t \Delta q_{xx} - (4Dr + \alpha) q_{xx} \\
    \partial_t q_{yy}&= -\frac{v}{4} \nabla \cdot \left[ \begin{pmatrix} -1 \\ 1\end{pmatrix} \boldsymbol{p} \right] + D_t \Delta q_{yy} - (4Dr + \alpha) q_{yy}.
\end{align}
By direct comparison with Eq~\eqref{eq:coefficients 2} and using the expression that we found for $\boldsymbol{p}$ we can show that 
\begin{equation}
    \underline{\underline{q}}(\boldsymbol{r},t) = \begin{pmatrix}
        \frac{1}{2}(a_2 + a_{-2}) &i(a_2 + a_{-2} ) \\ i(a_2 + a_{-2} ) & -\frac{1}{2}(a_2 + a_{-2})
    \end{pmatrix}.
\end{equation}
The coefficients we derived can therefore be mapped onto hydrodynamic fields with clearer physical meaning. 

Going at higher orders is not necessarily useful, since coefficients of order 3 and above are unclear in terms of physical meaning. 

\section{Detailed solution at order 2}
\label{appendix:math details}

\begin{comment}
The two poles are in the superior part of the complex plane and according to the Residue Theorem we then have  :
\begin{equation*}
    F(\boldsymbol{k},t) = 2\pi i \Biggl[Res[f(\omega),\omega_+]+Res[f(\omega,\omega_-)]\biggr]
\end{equation*}
With $ f(\omega) = \frac{1}{2\pi}e^{i\omega t} \frac{i\omega + D_tk^2+D_r}{(i\omega+D_tk^2)(i\omega+D_tk^2+D_r)+\frac{v^2k^2}{2}}$ and $ Res[f(\omega),\omega_\pm]=\lim_{\omega\to\omega_\pm} (\omega-\omega_\pm)f(\omega) $.
\\
The residues can be computed with the fact that $(i\omega+D_tk^2)(i\omega+D_tk^2+D_r)+\frac{v^2k^2}{2}=-(\omega-\omega_+)(\omega-\omega_-)$ and by taking the limit and we obtain :
\begin{equation*}
    Res[f(\omega),\omega_\pm] = \mp\frac{1}{2\pi}\frac{D_r\pm i\sqrt{2k^2v^2-D_r^2}}{2\sqrt{2k^2v^2-D_r^2}}e^{-\left(D_tk^2+\frac{D_r}{2}\mp\frac{i}{2}\sqrt{2k^2v^2-D_r^2}\right)t}
\end{equation*}
\end{comment}

This section details the calculation used to obtain the result for $F_2(k,t)$, the ISF obtained by truncating at order $N=2$. We start from 

\begin{equation*}
    F(\boldsymbol{k},t) = \int \frac{d\omega}{2\pi} e^{i\omega t} \widetilde{a}_{0,2}(\boldsymbol{k},\omega). 
   % & = \int \frac{d\omega}{2\pi} e^{i\omega t}\frac{2(i\omega +D_tk^2 +4D_r +\alpha)(i\omega +D_tk^2 +D_r +\alpha)+\frac{v^2k^2}{2}}{(i\omega +D_tk^2)[2(i\omega +D_tk^2 +4D_r +\alpha)(i\omega +D_tk^2 +D_r +\alpha)+\frac{v^2k^2}{2}]+v^2k^2(i\omega+D_tk^2+4D_r+\alpha)}
\end{equation*}

The denominator of $\widetilde{a}_{0,2}(\boldsymbol{k},\omega)$ is a polynomial of order 3 in $\omega$ that will be cancelled by three roots. Introducing the notations 
\begin{align*}
    a &= -8 \alpha^3 -60 \alpha^2D_r + 12\alpha D_r^2 + 280 D_r^3 +81D_r k^2 v^2 \\
    b &= 4 \alpha ^2 + 20\alpha D_r +52 D_r^2 - 9 k^2v^2
\end{align*}
we can find the roots to be 
\begin{align*}
    \omega_1 &= \frac{1}{6}i\bigg( (a+\sqrt{a^2-b^3})^{1/3} + \frac{(a+\sqrt{a^2-b^3})^{1/3}}{b} \\
    &\qquad +4\alpha +10D_r + 6D_t k^2\bigg) \\
    \omega_2 &= -\frac{1}{12}i \bigg( (1-i\sqrt{3})(a+\sqrt{a^2-b^3})^{1/3} \\ &+ (1+i\sqrt{3})\frac{(a+\sqrt{a^2-b^3})^{1/3}}{b} \\
    &\qquad - 8\alpha -20D_r - 12D_t k^2 \bigg) \\
    \omega_3 &= -\frac{1}{12}i \bigg( (1+i\sqrt{3})(a+\sqrt{a^2-b^3})^{1/3}\\&+ (1-i\sqrt{3})\frac{(a+\sqrt{a^2-b^3})^{1/3}}{b} \\
    &\qquad - 8\alpha -20D_r - 12D_t k^2 \bigg).
\end{align*}
These three poles are always in the superior part of the complex plane and therefore building a contour around it we get with the residue theorem 
\begin{equation}
\label{eq:residue order2}
    F_2(\boldsymbol{k},t) =  i \Biggl[\sum_{i=1}^3 \text{Res}[\widetilde{a}_{0,2}(\boldsymbol{k},\omega),\omega_i]\biggr].
\end{equation}
The denominator of $\widetilde{a}_{0,2}(\boldsymbol{k},\omega) $can be expressed in terms of these roots as 
\begin{align*}
    &(i\omega +D_tk^2) \\ 
    & \left[2(i\omega +D_tk^2 +4D_r +\alpha)(i\omega +D_tk^2 +D_r +\alpha) +\frac{v^2k^2}{2} \right] \\ &+v^2k^2(i\omega+D_tk^2+4D_r+\alpha) = -2i\prod_{i=1}^3 (\omega-\omega_i).
\end{align*}
The residues are then easily calculated by replacing the denominator, and introducing the numerator divided by 2 as $h(\omega) = (i\omega +D_tk^2 +4D_r +\alpha)(i\omega +D_tk^2 +D_r +\alpha)+\frac{v^2k^2}{4}$ we find the exact expression from the main text Eq.~\ref{eq:f2kt} 
\begin{align}
    F_2(\boldsymbol{k},t) = -\sum_{i=1}^3 \frac{h(\omega_i)e^{i \omega_i t}}{\prod_{j \neq i} (\omega_i - \omega_j)}.
\end{align}
As mentioned, this method is general and can be applied to solve at order 1 easily, and at order 3 to obtain a lengthier solution. At order 4 in the truncation however, the denominator will be of degree 5 in $\omega$ and it will not be possible to obtain general analytical solutions of its roots and thus of the integral anymore. %\tristan{Timothé et Pierre disent dans leur articles qu'on peut plus après l'ordre 3 pour cause de pas de solution générale de polynomes d'ordre 4 mais j'ai recheck et je suis quasiment sur qu'il y en a quoi, on veut pas les faires mais il y en a (ordre 5 c'est bien mort par contre)} 

\begin{figure*}[h!]
    \centering
    \includegraphics[width=0.99\linewidth]{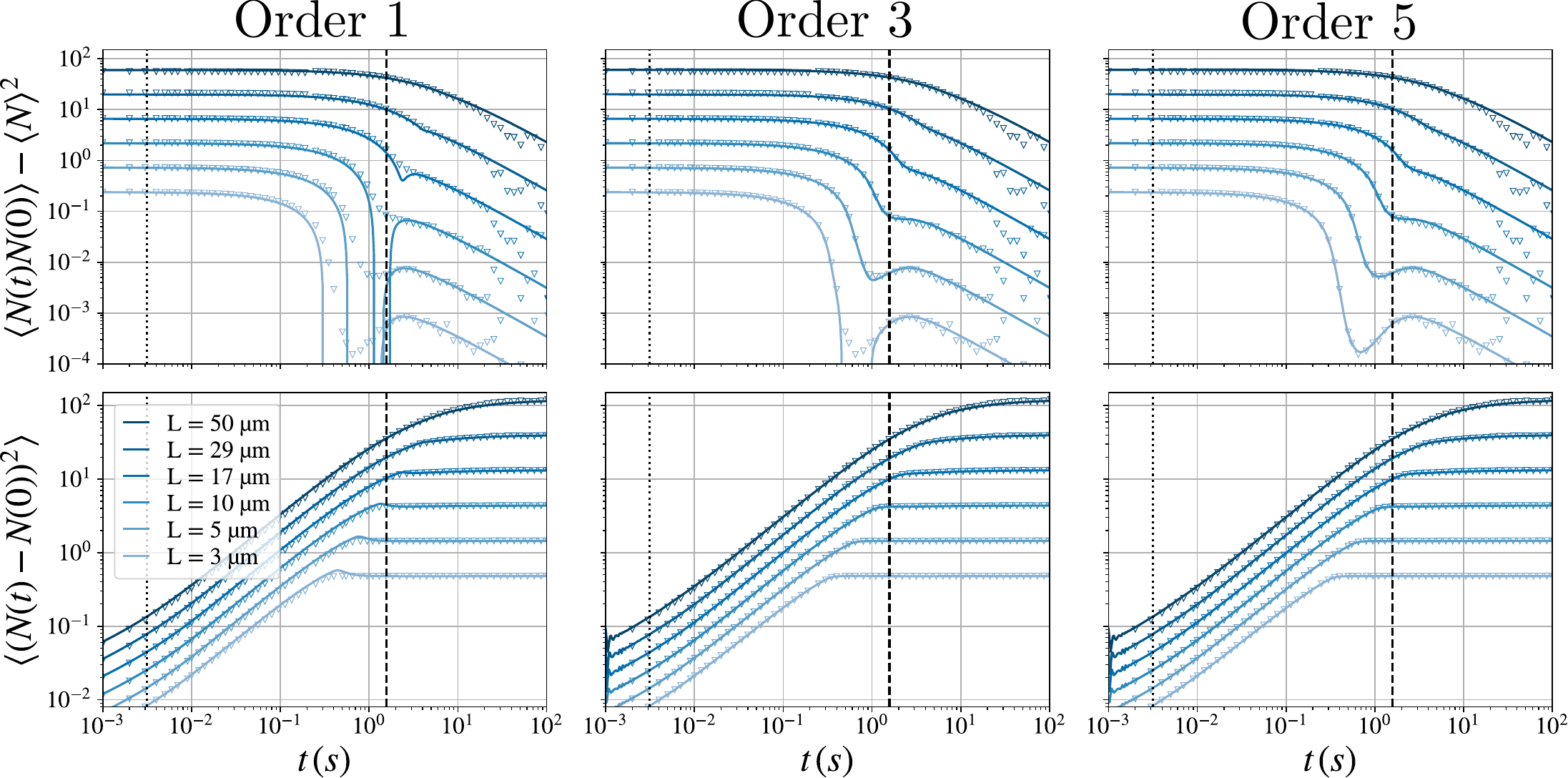}
    \caption{\textbf{Number correlations and NMSD for Active Brownian Particles at different Orders} in the truncation.}
    \label{fig:ABP Order Countoscope}
\end{figure*}

\begin{figure*}
    \centering
    \includegraphics[width=0.99\linewidth]{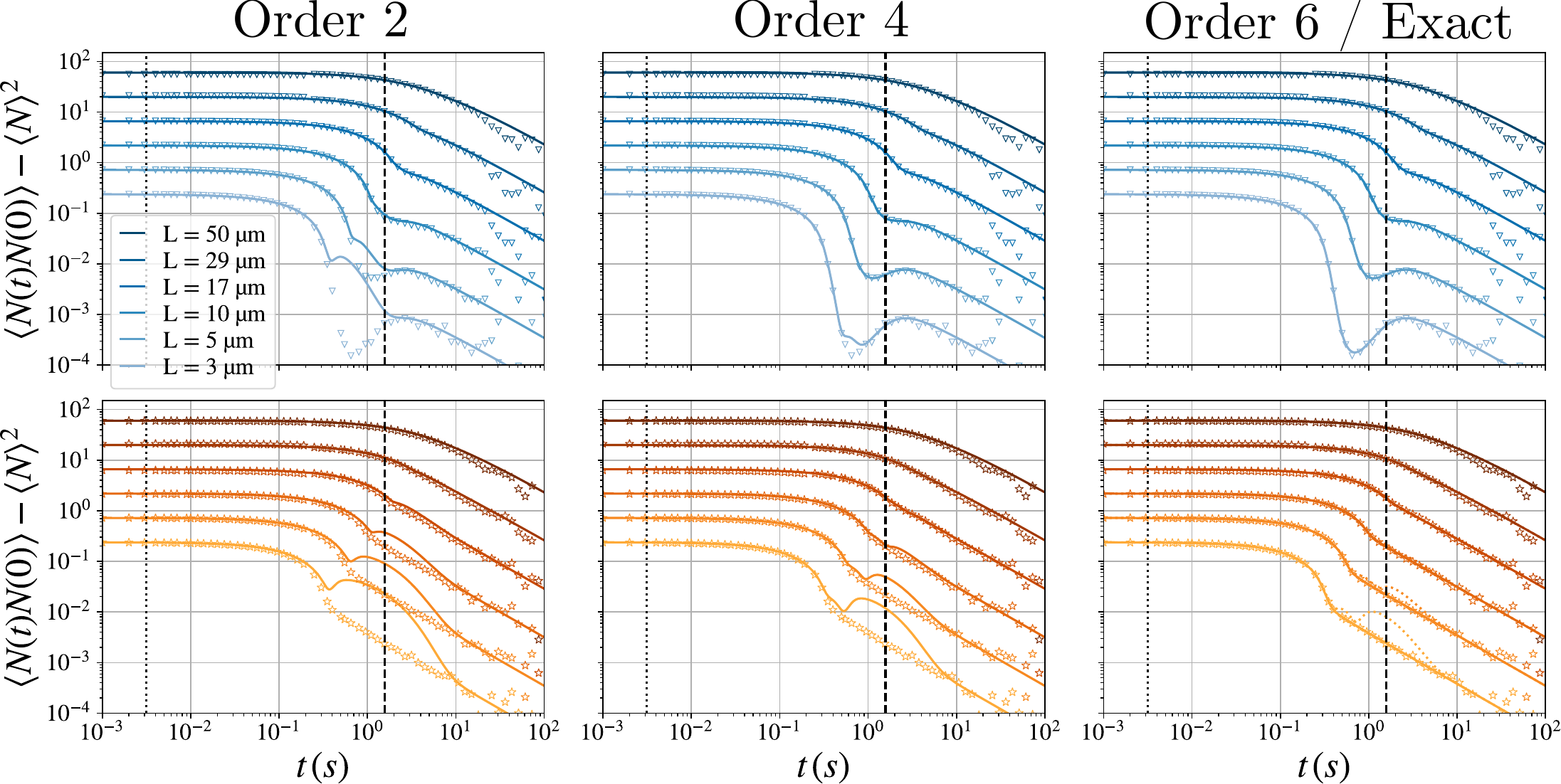}
    \caption{\textbf{Number Correlations} of Active Brownian Particles (top) and Run-and-tumble particles (bottom) at pair orders in the truncation. Orders are increasing from left to right, with the rightmost column for RTP showing the exact expression in full lines and the Order 6 result in dotted lines. Dynamic parameters are $v = 10~\unit{\mu m.s^{-1}} $, $D_t = 0.1~\unit{\mu m^2.s^{-1}} $, and $D_r = 1~\unit{s^{-1}}$.}
    \label{fig:Cnt pair orders}
\end{figure*}

\clearpage

\bibliography{biblio}

\end{document}